# Self-powered Solar Aerial Vehicles: Towards Infinite Endurance UAVs

Farbod Khoshnoud[a,c] , Ibrahim I. Esat[b], Clarence W. de Silva[c], Jason D. Rhodes[d], Alina A. Kiessling[d], Marco B. Quadrelli[d]

[a]*Department of Electromechanical Engineering Technology, College of Engineering, California State Polytechnic University, Pomona, CA, 93740 USA*
*E-mail: fkhoshnoud@cpp.edu*

[b]*Department of Mechanical and Aerospace Engineering, Brunel University London, Uxbridge UB8 3PH, UK*

*E-mail: Ibrahim.Esat@brunel.ac.uk*

[c]*Department of Mechanical Engineering, the University of British Columbia, Vancouver, BC V6T 1Z4, Canada*

*E-mail: desilva@mech.ubc.ca*

[d]*Jet Propulsion Laboratory, California Institute of Technology, 4800 Oak Grove Drive, Pasadena, CA 91109 USA*

*E-mails: jason.d.rhodes@jpl.nasa.gov; alina.a.kiessling@jpl.nasa.gov; marco.b.quadrelli@jpl.nasa.gov*

A self-powered scheme is explored for achieving long-endurance operation, with the use of solar power and buoyancy lift. The end goal is the capability of "infinite" endurance while complying with the UAV dynamics and the required control performance, maneuvering, and duty cycles. Nondimensional power terms related to the UAV power demand and solar energy input are determined in a framework of Optimal Uncertainty Quantification (OUQ). OUQ takes uncertainties and incomplete information in the dynamics and control, available solar energy, and the electric power demand of a solar UAV model into account, and provides an optimal solution for achieving a self-sustained system in terms of energy. Self-powered trajectory tracking, speed and control are discussed. Aerial vehicles of this class can overcome the flight time limitations of current electric UAVs, thereby meeting the needs of many applications. This paper serves as a reference in providing a generalized approach in design of self-powered solar electric multirotor UAVs.

## Nomenclature

| | | |
|---|---|---|
| $A$ | = | area |
| $A_{PV}$ | = | area associate with photovoltaic cells |
| $b_p$ | = | damping coefficient of the propellers |
| $B$ | = | body frame |
| $C_d$ | = | drag coefficient |
| $c_{eq}$ | = | equivalent electrical damping constant |
| $e$ | = | back electromotive force (emf) voltage |
| $f_i$ | = | thrust force associated with each propeller $i$ |
| $\mathbf{F}_B$ | = | force vector in body frame |
| $\mathbf{F}_{B,aero}$ | = | aerodynamic force vector in body frame |
| $\mathbf{F}_{B,thrust}$ | = | thrust force vector in body frame |
| $\mathbf{g}_I$ | = | gravity |
| $\mathbf{H}_I^B$ | = | rotation matrix |
| $i$ | = | propeller/motor number |
| $I$ | = | inertial frame |
| $\mathbf{I}$ | = | identity matrix |
| $I_0$ | = | reverse (dark) saturation current |

$I_{ai}$ = armature current
$I_{inertia}$ = moment of inertia
$I_d$ = diode current
$I_{PV}$ = photovoltaic cell circuit current applied to the load
$I_{SC}$ = short circuit current
$I_{SH}$ = current through the shunt resistor
$J_{mi}$ = moment of inertia of rotor i
$k$ = lift constant for a propeller
$k_B$ = Boltzmann's constant
$K_e$ = electric constant (voltage constant) of the motor
$K_t$ = torque constant of the motor
$k_{D,F,i}$ = derivative control gain associated with force $f_i$
$k_{I,F,i}$ = integral control gain associated with force $f_i$
$k_{P,F,i}$ = proportional control gain associated with force $f_i$
$k_{D,\boldsymbol{\alpha},i}$ = derivative control gain associated $\boldsymbol{\alpha}$ of propeller $i$
$k_{I,\boldsymbol{\alpha},i}$ = integral control gain associated $\boldsymbol{\alpha}$ of propeller $i$
$k_{P,\boldsymbol{\alpha},i}$ = proportional control gain associated $\boldsymbol{\alpha}$ of propeller $i$
$\mathbf{L}$ = transformation matrix; relationship between Euler-Angle rates and Body-Axis rates
$L_a$ = inductance
$l_i$ = distance from the center of the propeller i to the center of the gravity of the vehicle
$L_{aero}$ = aerodynamic moment about $x$ axis in body frame
$M_{aero}$ = aerodynamic moment about $y$ axis in body frame
$N_{aero}$ = aerodynamic moment about $z$ axis in body frame
$L_{thrust}$ = moment due to thrust forces about $x$ axis in body frame
$M_{thrust}$ = moment due to thrust forces about $y$ axis in body frame
$N_{thrust}$ = moment due to thrust forces about $z$ axis in body frame
$m$ = mass
$\mathbf{M}_B$ = moment vector in body coordinate
$\mathbf{M}_{B,aero}$ = aerodynamic moment vector in body frame
$\mathbf{M}_{B,thrust}$ = moment vector due to thrust forces in body frame
$n_i$ = ideality factor
$P_c$ = power consumption
$P_{ci}$ = power consumption by motor $i$
$P_g$ = generated power (by PV)
$P_{non}$ = nondimensional power
$p$ = angular velocity in body frame about $x_B$
$q$ = angular velocity in body frame about $y_B$
$r$ = angular velocity in body frame about $z_B$
$q_e$ = electron charge
$\boldsymbol{r}$ = position vector
$\boldsymbol{r}_B$ = position vector in body frame
$\boldsymbol{r}_1$ = position vector in an intermediate frame
$\boldsymbol{r}_I$ = position vector in inertial frame
$R_a$ = resistance
$R_s$ = resistance associated with voltage drop in the electrical contacts
$R_e$ = Reynolds number
$T_c$ = junction temperature
$T_i$ = torque generated by propeller i
$u$ = translational velocity in body frame along $x_B$
$v$ = translational velocity in body frame along $y_B$
$w$ = translational velocity in body frame along $z_B$
$v_a$ = voltage of the power source
$V_{PV}$ = photovoltaic voltage
$x_B$ = $x$ axis in body frame

$y_B$ = $y$ axis in body frame
$z_B$ = $z$ axis in body frame
$\boldsymbol{v}_B$ = velocity vector in body frame
$z_I$ = $z$ axis in inertial frame
$x_I$ = $x$ axis in inertial frame
$y_I$ = $y$ axis in inertial frame
$z_I$ = $z$ axis in inertial frame
$X_{aero}$ = aerodynamic force along $x$ axis in body frame
$Y_{aero}$ = aerodynamic force along $y$ axis in body frame
$Z_{aero}$ = aerodynamic force along $z$ axis in body frame
$X_{thrust}$ = thrust force along $x$ axis in body frame
$Y_{thrust}$ = thrust force along $y$ axis in body frame
$Z_{thrust}$ = thrust force along $z$ axis in body frame
$X$ = deterministic variables
$\boldsymbol{\alpha}_i$ = angle vector associated with thrust vector with respect to body frame axis for each propeller $i$
$\alpha$ = drive/regenerative mode a motor
$\boldsymbol{\beta}_i$ = direction angles with respect to the body frame
$\eta$ = efficiency
$\theta$ = pitch rotation angle measured in an intermediate frame
$\dot{\boldsymbol{\Theta}}$ = Euler angle rate vector
$\mu_j$ = probability measure
$\rho$ = density of air
$\phi$ = roll rotation angle measured in an intermediate frame
$\psi$ = yaw rotation angle measured in the inertial frame
$\boldsymbol{\omega}_B$ = angular velocity vector
$\omega_{pi}$ = angular velocity of the $i^{th}$ propeller
$\mathcal{X}$ = stochastic variables
CG = center of gravity
CV = center of volume
ESC = Electronics Speed Controllers
LEMV = Long Endurance Multi-Intelligence Vehicle
LTA = Lighter-than-air
MAAT = Multibody Advanced Airship for Transport
OUQ = Optimal Uncertainty Quantification
PV = photovoltaic
PWM = Pulse Width Modulation
UAV = Unmanned Aerial Vehicles

# I. Introduction

Autonomous aerial robots (Aerobots) have the potential to make significant positive to both commercial and public service aviation, with land, sea, air and space applications [1]-[4]. Notably, much interest is found in multirotor configurations. Multirotor configurations normally suffer from low rotor/vehicle aerodynamic efficiencies both in hover and loitering flight and in cruise [5]. Thus, the limited flight time is a main challenge in the area of multirotor electric Unmanned Aerial Vehicles (UAVs) with vertical take-off and landing and station-keeping capabilities. Once long endurance electric UAVs are realizable, the fast growing demand of such systems can be addressed. Applications are found in the areas of safety, security, surveillance, emergency, delivery (e.g., [6]), search and rescue (e.g., [7]-[9]), robotic and space exploration (e.g., [10]-[12]), maintenance, traffic, monitoring, transport, telehealth, and beaming internet. Multirotor configurations represent a potentially rich design space that has only just begun to be explored [5]. Novel solar-powered multirotor electric aerial vehicles ([13]-[21]) have been developed to address the flight time limitation and energy consumption towards achieving self-powered and energy independent systems ([13]-[16], [22]). There is a rich literature available in the area of solar-powered unmanned aerial vehicles. However, only few references are given here in the following literature review as an introductory overview, which refers to the importance of the topic, and highlights the contributions of this paper.

Interest in development and use of long duration, lighter-than-air (LTA) vehicles for various applications has grown considerably in recent years, especially those with abilities of station-keeping, autonomy, and long endurance operation [23]. Airships gain their altitude through buoyancy and/or propulsion systems, with enclosed lightweight gas such as hydrogen or helium. Capability of hovering, and operating for a long time at varying altitudes without refueling at relatively low costs are potential advantages of airship over aircraft [24].

Platforms that utilize LTA technologies such as aerostats (buoyant craft tethered to the ground), and airships (buoyant craft that are free-flying) may hold the potential for significantly increasing capacity in the areas of persistent Intelligence, Surveillance, and Reconnaissance (ISR) and communications, as well as lowering the costs of transporting cargo over long distances, without the need for aircraft runways. The aerostat system provides a link to the information and intelligence network that is going to be reliable 24 hours a day 7 days a week for long periods of time [25]. By using on-board integrated optical sensors, thermal imaging and radar, near-total awareness of the surface below can be achieved by such aerial vehicles. Sea surveillance is attainable from an elevated prospective from above by gathering information for detecting small high-speed aerial (e.g. low flying aircraft) and maritime targets at distances of more than 60 nautical miles, with 360° view, for more than two weeks at a time. "Aerostat is affordable, reliable, dependable, and absolutely efficient" [25].

Department of Defense's pursuit of aerostats and airships is mostly due to the ability of these platforms to loiter for a longer period of time in comparison with fixed-wing unmanned aircraft, which makes them suitable for supporting the ISR missions. The aerostat and airship on-station endurance time is typically greater than that of fixed-wing unmanned aircraft. For example, the Long Endurance Multi-Intelligence Vehicle (LEMV) United States Army Airship is expected to stay on station for at least 16 days. The amount of time on station is greatly dependent on how often the aerostat or airship needs to be topped off with additional helium/LTA gas. Fixed-wing unmanned aircraft can normally stay on station for shorter duration of time (For example, 6 hours for the Shadow aircraft, and 40 hours for the Sky Warrior) [26].

New developments of fixed wing aircraft such as Solar Impulse, or NASA's Helios can potentially offer longer duration of flight. Aircraft do not offer station-keeping and low speed flight capabilities. Furthermore, airship do not require runway and the associated infrastructure for takeoff and landing. Another important note on the advantage of airships over other air vehicles is that, in case of failure of the system during the flight, the airship inherently acts like an airbag cushion. It will not damage any object or injure a person in a crash, which is a great advantage particularly when flying in urban areas. Moreover, if the helium envelope is torn or punctured, leakage of helium from the envelope is a gradual process leading to a relatively safe and gentle landing.

Manned airships such as Airlander 10 [27], the Lockheed Martin Hybrid Airship [28], and Solarship [29] are examples of recent advancements in large cargo airship technologies. Space exploration buoyant robotic vehicles (aerobots) have been proposed for various missions in extreme planetary environments such as exploration of planets and strategic surveying of moons with an atmosphere, such as Venus, Mars and Titan [1]-[3], [12]. Aerobots offer "modest power requirements, extended mission duration and long traverse capabilities, and the ability to transport and deploy scientific instruments and in-situ laboratory facilities over vast distances" [30]-[32].

Multibody Advanced Airship for Transport (MAAT) (e.g. [14], [16], [33]-[37]) proposes the application of airships for passenger and cargo transport while the airship is composed of a multibody (multi airship) system. In this concept, smaller airships transport passengers/cargo to other airship carriers and to the ground. An analogy to the exchange of small airships in between larger airship carriers can be made to a train station or airport transit when passengers are transferred to other trains/aircrafts to reach to a desired final destination. The energy supply of the entire system, including control and propulsion systems, is produced by an integrated solar-fuel cell system. The electric energy for MAAT operation is supplied by a photovoltaic (PV) system, during daytime, and hydrogen based fuel cells, during night, using the hydrogen produced by PVs. A photovoltaic generator that operates an electrolyzer, produces hydrogen during the day. At night, a proton exchange membrane fuel cell uses the hydrogen to supply electrical power. The produced energy can provide 24-hour operation for MAAT with no impact on greenhouse gas emissions.

In 2013 the Keck Institute for Space Studies (KISS [51]) sponsored a yearlong study entitled "Airships: A New Horizon for Science" [52], with the purpose of enabling science and commercial applications. This workshop brought together people from the astrophysics community, the planetary science community, the Earth science community, and potential commercial suppliers of stratospheric airships. The study entailed two-week long workshops and developed space science cases for airships at a range of altitudes. The final report from that study [23] presented a strong case for stratospheric airships as science platforms. The report also identified Earth science as a key driver for stratospheric airships capability and enumerated multiple commercial applications in remote site monitoring and communication.

The KISS study recommended that in order to incentivize the commercial development of persistent (weeks or months) stratospheric airships a monetary prize be attached to the effort. One obvious mechanism for this would be a NASA Centennial Challenge [53] program in which NASA provides cash prizes to organizations that meet well defined challenge goals. This prompted the development of a challenge concept called the "Airships 20-20-20 Challenge; [54]". This would be a multi tiered challenge in which the first team to fly an airship at 20 km for 20 hours (one diurnal cycle) with a 20 kg payload would receive a modest cash prize. The second stage of the challenge would be for a team to fly an airship for 200 hours at 20 km with a 200 kg payload. This would result in a more substantial cash prize. These would both be stepping stones to larger airships with ~1000 kg class payloads flying for weeks or months; this would be an extremely compelling platform for multiple areas of astrophysical science. During the challenge development, it became clear that there are also many compelling Earth Science investigations that could be done at both levels of the challenge (that is, flying 20 kg for 20 hours or flying 200 kg for 200 hours, both at 20 km altitude). NASA headquarters is currently in the process of deciding whether to fund the Airships 20-20-20 Challenge.

Stability and control for navigation, stable station-keeping and maneuvering for free flying airships are challenging due to buoyancy, size and shape. The technology associated with the control systems and stability of airships has not been matured as much as for fixed wing aircraft, helicopter, and multirotor drones. Control, energy and propulsion systems of airships have been studied by many researchers and a brief overview of the related research in this area is presented below.

Various modelling techniques have been implemented for simulating and analysis of airship dynamics, stability and control [2], [3], [38]. For instance, a six-degree-of-freedom model of a rigid airship can cover a complete flight envelope in studying the airship dynamic behavior. The simulation results from such model can be used at various stages of design process [39]. A comprehensive dynamic model includes aerodynamic lift and drag, vectored thrust, added mass effects, and accelerations due to mass-flow rate, wind rates, and Earth rotation in the equations of motion of the airship [40]. Control of altitude in large airships is normally achieved by varying the amount of air filled in a ballonet inside the airship hull, which adjusts the buoyancy force. The amount of ambient gas filled in ballonets inside airship hulls can be controlled resulting in change of altitude [12], while fins and vertical thrust propellers, or thrust vectoring can also control vertical motion of the airships. Hybrid airships take advantage of aerodynamic lift using wings or wing shape hulls in addition to buoyancy control system or thrust of propellers [41]. Autonomous station-keeping is one of the key challenges of airships, and significant thrust and control power is required to station keep in turbulence [42]. The dynamic model and control system should take into account the nonlinear behavior of the airship in presence of wind disturbances while propelling with limited number of actuators (e.g. two independent propellers without side thrusters) [43]. In addition to obtaining analytical models of airships (e.g. using the 6DOF equations of motion), Artificial Neural Networks (ANN) can learn from flight test data of robotic airships following a trajectory obtained from human manual control ([44], [45]). For specific cases where airships operating at high-altitudes, various environmental conditions such as near-AM0 spectrum, high voltage Paschen discharges, thermal extremes, thermal shock, and atomic oxygen exposure are particularly important and should be taken into account [46]. It has been shown that sufficient power can be generated for operations and energy demand of an airship at various altitudes using photovoltaic solar cells, which mainly depends on the solar cell dimensions and efficiency, and orientation of cells during the flight [47].

Brunel Solar UAVs ([13], [15], [17]-[21]) are novel systems designed to realize long endurance flight and autonomy. The multirotor configurations along with solar power, autonomy, and small size buoyancy hull make them unique systems.

This paper explores conditions and scenarios for developing Self-powered Solar UAVs as energy independent vehicles in accordance to dynamics, control system and accessible solar energy of the vehicles. The concept of Self-powered Dynamic System ([13]-[16], [22]) in the framework of Optimal Uncertainty Quantification ([48], [13], [15]) is applied here. OUQ [48] takes uncertainties and incomplete information in the dynamics, control, available energy, and electrical power demand of the system into account, and provide optimal solutions in attaining self-powered control and maneuvering with sustainable energy independent condition. This paper is organized as follows. Section II describes Solar-powered Multirotor Aerial Vehicle. Section III deals with the Dynamics of Multirotor Aerial Vehicles. Section IV describes Self-powered Dynamic Systems. Section V summarizes the elements of Optimal Uncertainty Quantification. Section VI addresses Solar-powered Speed. Section VII covers self-powered trajectory tracking. Section VII addresses self-powered dynamics and control. Section IX deals with Enabling Science and Commercial Applications, and Section X concludes the paper.

## II.Solar-powered Multirotor Aerial Vehicles

Three solar-powered vehicles known as Brunel Quadrotor, Octorotor, and Trirotor solar powered UAVs ([13], [15], [17]-[21]) have been developed for achieving self-sustained systems in terms of energy in the framework of ‘Self-powered Dynamic Systems’ ([13]-[16], [22]). This framework seeks to realize dynamic systems powered by their excessive kinetic energy, renewable energy sources, or a combination of both. Self-powering through solar UAVs is discussed now. In particular Brunel solar UAVs are introduced. Further information on the specifications is available in references [17]-[19]. The hull sizes of the quadrotor system in Figure 1, octorotor in Figure 2, and trirotor in Figure 3 are 3×2×1 m, 2.5 m in diameter, and 2.2 m length, 1.8 m width and 1.1m height, respectively. The vehicles are equipped with flight control systems (e.g. Pixhawk flight controller hardware), and sensors such as accelerometers, gyroscopes, global positioning systems, magnetometers, barometers, etc. These UAVs are powered, fully or partially (they are still under development and further investigation is required), by solar energy while taking advantage of buoyancy lift for maintaining the vehicles aloft with no, or reduced, thrust force for lift. Monocrystalline photovoltaic (PV) cells, organic PV cells, and amorphous silicon PVs supply the energy in the quadrotor, octocopter, and the tricopter UAVs, respectively. PV cells charge the batteries, as high power density source, and the trirotor is additionally equipped with a fuel cell system, as a high energy density source. The quadrotor, octorotor, and trirotor configurations exhibit different dynamics behavior in terms of control and stability aspects, aerodynamics, agility, performance, and energy consumption, due to their sizes and geometries of the blimps. The power demand for such UVAs is in the range of 100’s of Watts up to Kilowatts range. The dynamics of multirotor UAVs is presented in the next section, followed by the study of achieving self-powered condition.

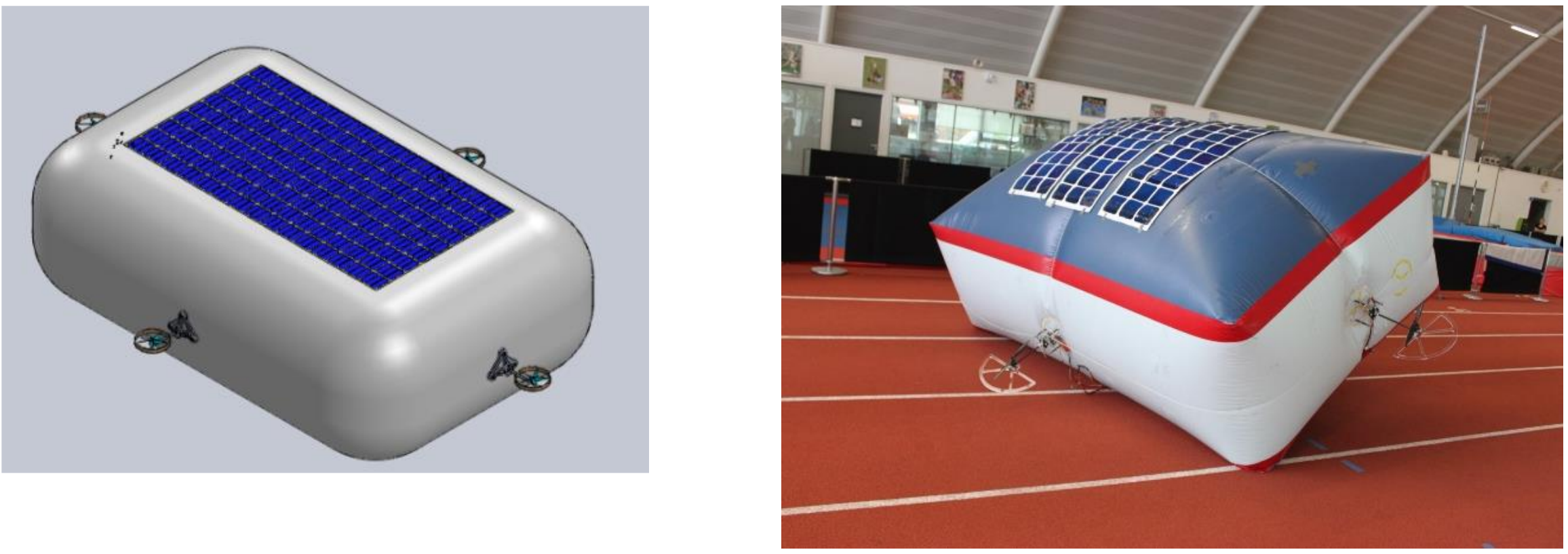

Figure 1. The Solar-powered Quadrotor UAV.

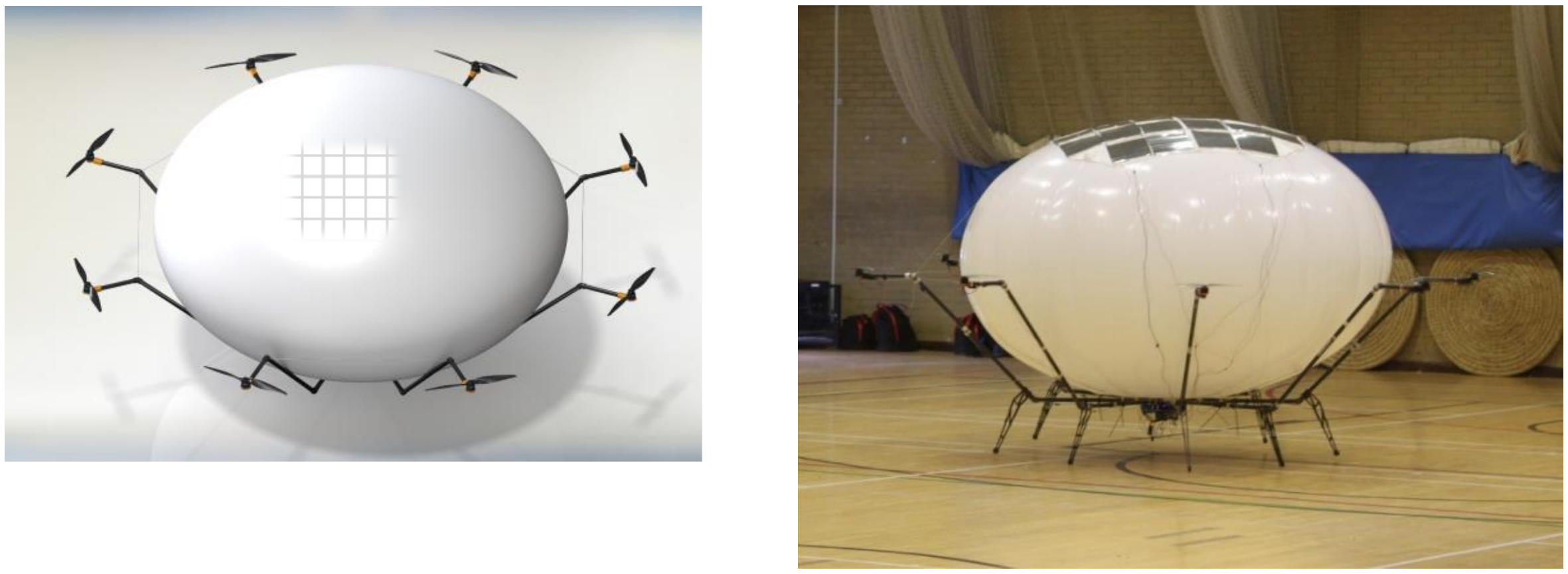

Figure 2. The Solar-powered Octorotor UAV.

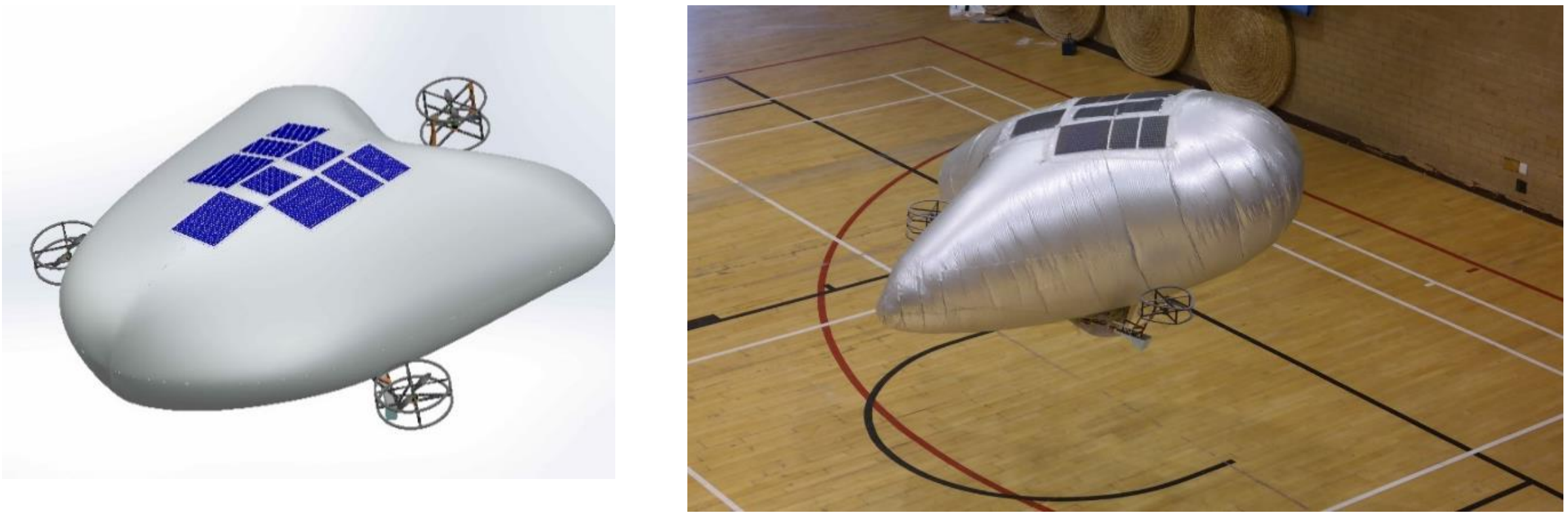

Figure 3. The Solar-powered Trirotor UAV.

## III.Dynamics of Multirotor Aerial Vehicles

Assumptions that are made in deriving the mathematical model of the flight dynamics of lighter-than-air (LTA) UAVs include: a) the UAVs virtual (added) mass and inertia due to the large volume of air displaced by the airship; b) the vehicle motion is referenced to a system of orthogonal axes fixed to the vehicle whose origin is the center of volume (CV) and is assumed to coincide with the gross center of buoyancy (CB); (c) the vehicle is a rigid body (i.e., aero-elastic effects are neglected). The orientation of the axes with respect to an Earth-fixed frame can be defined by Euler angles. The airship linear velocities and angular velocities (referred to as the roll, pitch and yaw rates) are determined in this coordinate system. Virtual mass and moments of inertia of the airship are considerably large in view of the large volume of displaced air. The geometry of the airship is assumed to be symmetric about one of its vertical planes and the CV and the center of gravity (CG) are located on that plane. A feedback control system ensures the stability of the system, and achieves capabilities for path following and trajectory tracking, as discussed in the next sections.

The equations of motion for multirotor aerial vehicles are presented below. The moment of inertia matrix of the vehicle can be written as

$$I_{inertia} = \begin{bmatrix} I_{xx} & -I_{xy} & -I_{xz} \\ -I_{yx} & I_{yy} & -I_{yz} \\ -I_{zx} & -I_{zy} & I_{zz} \end{bmatrix}$$

For principal axes, it introduces

$$I_{inertia} = \begin{bmatrix} I_{xx} & 0 & 0 \\ 0 & I_{yy} & 0 \\ 0 & 0 & I_{zz} \end{bmatrix}$$

The Euler angles for the sequence of rotations from inertial frame, $I$, to body frame, $B$, can be given by yaw, then pitch, and then roll, with yaw rotation $\psi$ about $z_I$, in Inertial Frame, or in mathematical form $\boldsymbol{r}_1 = \mathbf{H}_I^1 \boldsymbol{r}_I$, pitch rotation $\theta$ about $y_1$, in Intermediate Frame, or $\boldsymbol{r}_2 = \mathbf{H}_1^2 \boldsymbol{r}_1$, and roll rotation $\phi$ about $x_2$, in the Body Frame, as $\boldsymbol{r}_B = \mathbf{H}_2^B(\phi)\boldsymbol{r}_2$.

The three-angle rotation matrix is the product of 3 single-angle rotation matrices as

$$\mathbf{H}_I^B = \mathbf{H}_2^B(\phi)\mathbf{H}_1^2(\theta)\mathbf{H}_I^1(\psi)$$

and $[\mathbf{H}_I^B]^{-1} = [\mathbf{H}_I^B]^T = \mathbf{H}_B^I$; and $\mathbf{H}_B^I\mathbf{H}_I^B = \mathbf{H}_I^B\mathbf{H}_B^I = \mathbf{I}$.

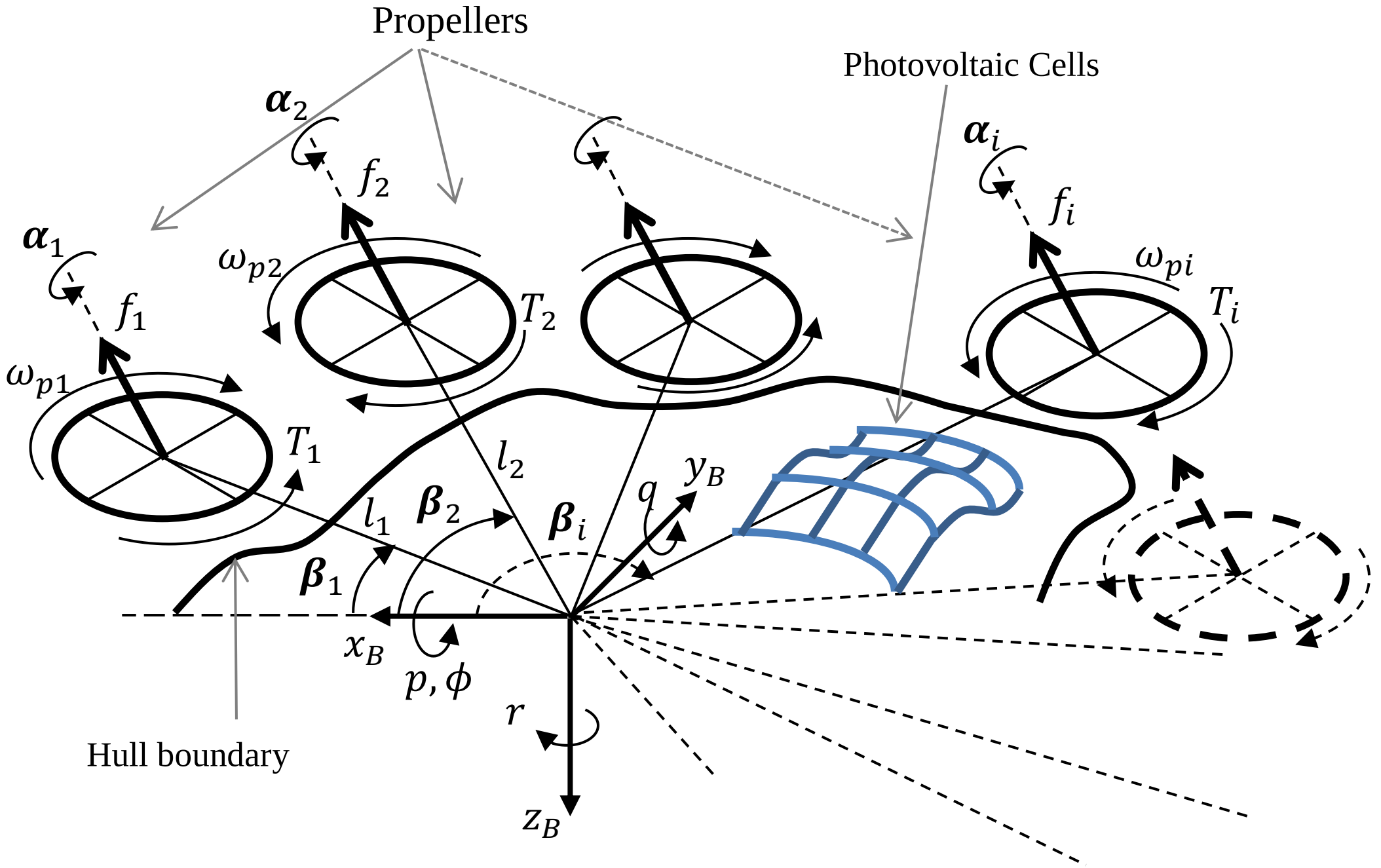

Figure 4. Schematic of a generalized multirotor aerial system.

The relationship between Euler-Angle rates and Body-Axis rates is expressed as

$$\begin{bmatrix} p \\ q \\ r \end{bmatrix} = \begin{bmatrix} 1 & 0 & -\sin\theta \\ 0 & \cos\phi & \sin\phi\cos\theta \\ 0 & -\sin\phi & \cos\phi\cos\theta \end{bmatrix} \begin{bmatrix} \dot{\phi} \\ \dot{\theta} \\ \dot{\psi} \end{bmatrix} = \mathbf{L}_I^B \dot{\Theta}$$

Or

$$\begin{bmatrix}\dot\phi\\ \dot\theta\\ \dot\psi\end{bmatrix} = \begin{bmatrix}1 & \sin\phi\tan\theta & \cos\phi\tan\theta\\ 0 & \cos\phi & -\sin\phi\\ 0 & \sin\phi/\cos\theta & \cos\phi/\cos\theta\end{bmatrix}\begin{bmatrix}p\\ q\\ r\end{bmatrix} = \mathbf{L}_{\boldsymbol{B}}^{\boldsymbol{I}}\omega_B$$

where $\dot\psi$ is measured in the inertial frame, $\dot\theta$ is measured in the intermediate frame, and $\dot\phi$ is measured in the body frame. $\dot{\mathbf{\Theta}} = \begin{bmatrix}\dot\phi\\ \dot\theta\\ \dot\psi\end{bmatrix}$ denotes the Euler angle rate vector, and $\boldsymbol{\omega}_B = \begin{bmatrix}\omega_x\\ \omega_y\\ \omega_z\end{bmatrix}_B = \begin{bmatrix}p\\ q\\ r\end{bmatrix}$ is the angular velocity vector in body frame.

The rate of change of translational position is obtained by

$$\dot{\mathbf{r}}_I(t) = \mathbf{H}_{\boldsymbol{B}}^{\boldsymbol{I}}(t)\boldsymbol{v}_{\boldsymbol{B}}(t)$$

where $\mathbf{r}_I = \begin{bmatrix}x\\ y\\ z\end{bmatrix}_I$ is the translational position, and $\boldsymbol{v}_{\boldsymbol{B}} = \begin{bmatrix}u\\ v\\ w\end{bmatrix}$ is the translational velocity in body frame.

The rate of change of angular position is

$$\dot{\mathbf{\Theta}}(t) = \mathbf{L}_{\boldsymbol{B}}^{\boldsymbol{I}}(t)\boldsymbol{\omega}_B(t)$$

The equation of motion can be written in terms of the rate of change of translational velocity as

$$\dot{\boldsymbol{v}}_{\boldsymbol{B}}(t) = \frac{1}{m(t)}\mathbf{F}_B(t) + \mathbf{H}_I^B(t)\mathbf{g}_I - \widetilde{\boldsymbol{\omega}}_B(t)\boldsymbol{v}_{\boldsymbol{B}}(t) \tag{1}$$

and the rate of change of angular velocity is

$$\dot{\boldsymbol{\omega}}_B(t) = \boldsymbol{I}_B^{-1}(t)[\mathbf{M}_B(t) - \widetilde{\boldsymbol{\omega}}_B(t)\boldsymbol{I}_B(t)\boldsymbol{\omega}_B(t)] \tag{2}$$

where $\widetilde{\boldsymbol{\omega}}$ is the cross product equivalent matrix of $\boldsymbol{\omega}$ given by

$$\widetilde{\boldsymbol{\omega}} = \begin{bmatrix}0 & -\omega_z & \omega_y\\ \omega_z & 0 & -\omega_x\\ -\omega_y & \omega_x & 0\end{bmatrix}$$

and $\dot{\mathbf{H}}_I^B = -\widetilde{\boldsymbol{\omega}}_B\mathbf{H}_I^B$; $\dot{\mathbf{H}}_B^I = \widetilde{\boldsymbol{\omega}}_I\mathbf{H}_B^I$.

The external forces include the aerodynamic and thrust forces in body frame as

$$\mathbf{F}_B = \mathbf{F}_{B,aero} + \mathbf{F}_{B,thrust} = \begin{bmatrix}X_{aero} + X_{thrust}\\ Y_{aero} + Y_{thrust}\\ Z_{aero} + Z_{thrust}\end{bmatrix}_B \tag{3}$$

and the applied moments in body frame are

$$\mathbf{M}_B = \mathbf{M}_{B,aero} + \mathbf{M}_{B,thrust} = \begin{bmatrix}L_{aero} + L_{thrust}\\ M_{aero} + M_{thrust}\\ N_{aero} + N_{thrust}\end{bmatrix}_B \tag{4}$$

The thrust force $f_i$ associated with each propeller $i$ is related to the angular velocity, $\omega_{pi}$ of the $i^{\text{th}}$ propeller, by

$$f_i = k\omega_{pi}^2 \tag{5}$$

where $k$ is the lift constant. For tilt rotor capability, each propeller $i$ has a thrust vector along $\boldsymbol{\alpha}_i$ vector with components of $\alpha_{ix}$, $\alpha_{iy}$ , and $\alpha_{iz}$ along body frame axes $x_B$, $y_B$, and $z_B$, respectively. Therefore, the thrust force components in Equation (3) can be written as $(X_{thrust})_B = \sum_{i=1}^n f_i \cos\alpha_{ix}$, $(Y_{thrust})_B = \sum_{i=1}^n f_i \cos\alpha_{iy}$, and $(Z_{thrust})_B = \sum_{i=1}^n f_i \cos\alpha_{iz}$, where $n$ is the number of propellers. The aerodynamic forces can be given by $(X_{aero})_B = {}^1\!/_2\, C_d\rho A u^2$, $(Y_{aero})_B = {}^1\!/_2\, C_d\rho A v^2$, and $(Z_{aero})_B = {}^1\!/_2\, C_d\rho A w^2$, where $A$ is the frontal area of the vehicle.

If $l_i$ is the distance from the center of the propeller $i$ to the center of mass of the vehicle, and $\boldsymbol{\beta}_i$ the corresponding direction angles with respect to the body frame, the moments due to the propeller thrust forces can be written as

$$\begin{aligned}(L_{thrust})_B = \sum_{i=1}^n &\left[\left(f_i \cos\alpha_{iy} \times l_i\cos\beta_{iz}\right)\right.\\ &\left. + \left(f_i \cos\alpha_{iz} \times l_i\cos\beta_{iy}\right)\right]\\ &+ \sum_1^n T_i \cos\alpha_{ix}\end{aligned}$$

$$\begin{aligned}(M_{thrust})_B = \sum_{i=1}^n &[(f_i \cos\alpha_{ix} \times l_i\cos\beta_{iz})\\ &+ (f_i \cos\alpha_{iz} \times l_i\cos\beta_{ix})]\\ &+ \sum_1^n T_i \cos\alpha_{iy}\end{aligned}$$

$$\begin{aligned}(N_{thrust})_B = \sum_{i=1}^n &\left[\left(f_i \cos\alpha_{ix} \times l_i\cos\beta_{iy}\right)\right.\\ &\left. + \left(f_i \cos\alpha_{iy} \times l_i\cos\beta_{ix}\right)\right]\\ &+ \sum_1^n T_i \cos\alpha_{iz}\end{aligned}$$

where $T_i$ is the torque generated by propeller $i$.

The mechanical power consumption for a desired motion (e.g. associated with a trajectory reference) can be calculated in the inertial frame, $I$, as

$$P_c = \mathbf{F}_I.\boldsymbol{v}_I(t) + \mathbf{M}_I.\boldsymbol{\omega}_I(t) \tag{6}$$

The self-powered analysis of the system using an electromechanical model of the vehicle is discussed in the next section.

## IV. Self-powered Dynamic Systems

The concept of self-powered dynamic systems is used in the development of fully self-sustained energy independent devices and vehicles. They are capable of using renewable energy resources, or harvesting the excessive kinetic energy in the system using regenerative techniques, or a combination of the two ([13]-[16], [22]). In the present context self-powered dynamic systems are associated with the solar-powered UAVs (Brunel Quadrotor, Octorotor, and Trirotor UAVs [13], [15], [17]-[21]). It is important here to investigate the energy consumption of the UAV as a function of the duty cycle, and provide a reference and approach to identify the self-powered conditions in terms of inputs, outputs, and parameters of the system.

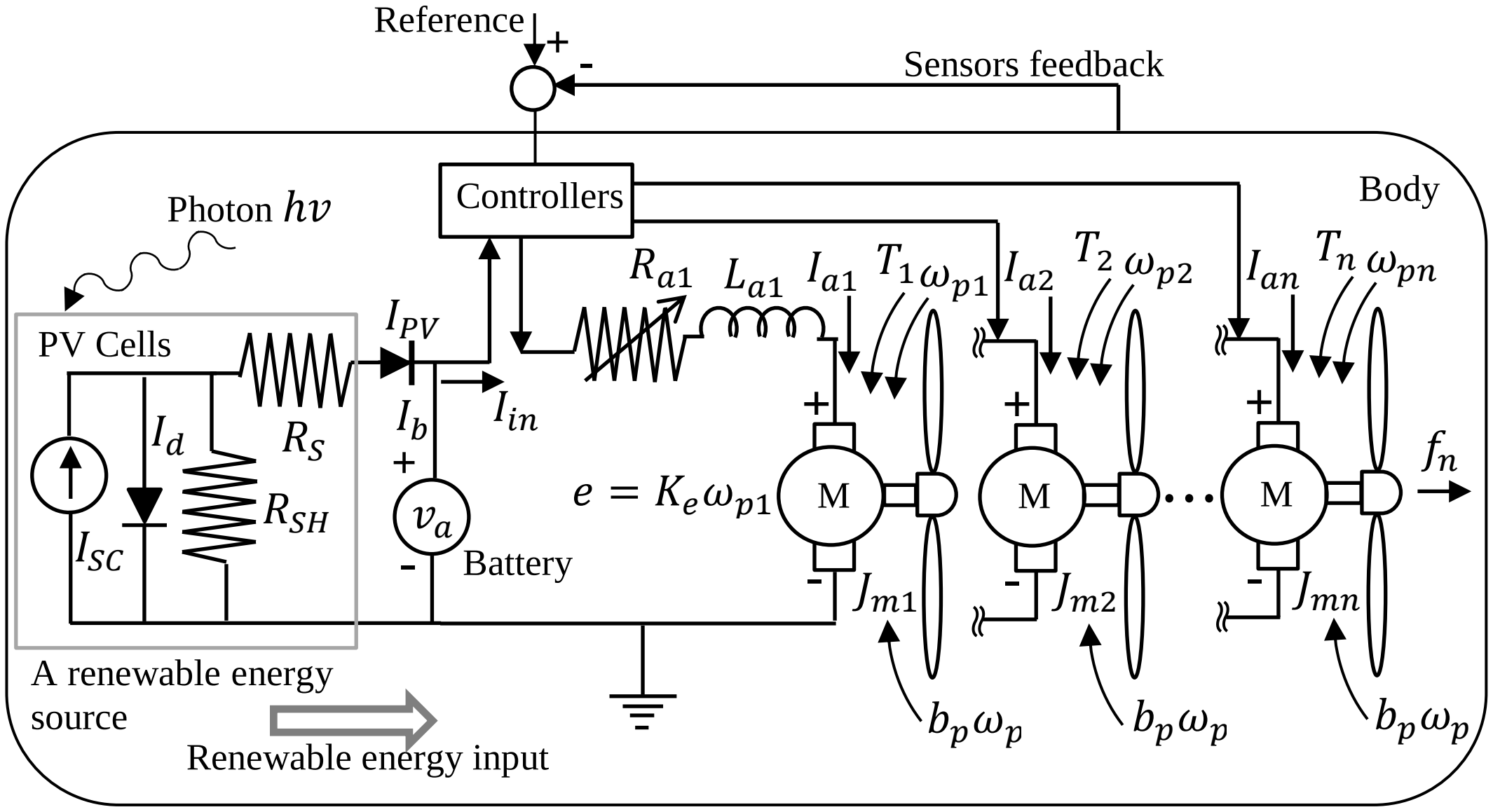

Figure 5. Electromechanical representation of the solar-powered vehicle.

The equation of motion of the vehicle are given by Equation (1), the Rate of change of translational velocity, and equation (2), the rate of change of angular velocity. The total power consumption, $P_c$, is given by Equation (6). If the power consumption by each electric motor $i$ (Figure 5) is $P_{ci}$ then

$$P_c = \sum_{i=1}^{n} P_{ci} \tag{7}$$

where $n$ is the total number of motors. The equation of motion for a motor can be written as

$$J_{mi}\dot{\omega}_{pi} + b_p\omega_{pi} = T_i = K_t I_{ai} \tag{8}$$

where $J_{mi}$ is the moment of inertia of rotor $i$, and $b_p$ is the damping coefficient of the propellers due to aerodynamic forces. The applied torque to the rotor, $T_i$, in terms of the armature current, $I_{ai}$, for DC linear motors can be given by

$$T_i = K_t\, I_{ai} \tag{9}$$

where $K_t$ is the torque constant of the motor. The back electromotive force (emf) voltage, $e$, in terms of the angular velocity of the shaft, $\omega_{pi}$, is given by

$$e = K_e\omega_{pi} \tag{10}$$

where $K_e$ is the electric constant (voltage constant) of the motor.

The equation for the electric circuit of the motor can be expressed by

$$L_a\frac{dI_{ai}}{dt} + R_a I_{ai} = v_a - K_e\omega_{pi} \tag{11}$$

where $v_a$ is the voltage of the power source, and $R_a$ is the resistance (Figure 5). If the effect of inductance, $L_a$, is negligible (Note: This is in fact the "leakage" inductance, which is relatively small in a good motor) compared to the effect of the resistance, then

$$R_a I_{ai} = v_a - K_e\omega_{pi} \tag{12}$$

For a DC motor, $K_t = K_e$ in consistent SI units. In order to generate the torque, $T_i$, the power consumed by the voltage source can be determined by substituting Equations (10) and (9) into (12) as

$$v_a = R_a\frac{T_i}{K_t} + k_e\omega_{pi} \tag{13}$$

The required torque can be calculated by rearranging Equation (13) as

$$T_i = K_t\frac{v_a - k_e\omega_{pi}}{R_a} \tag{14}$$

The consumed/required power can be obtained from Equations (9) and (13) as

$$P_{ci} = v_a I_{ai} = \left(R_a \frac{T_i}{K_t} + k_e \omega_{pi}\right) \frac{T_i}{K_t} \quad (15)$$

By taking into account the power received from the photovoltaic cells (PV cells), the total power consumption can be given by (when $K_t = K_e$)

$$P_{ci} = v_a I_b = v_a(\alpha I_{ai} - I_{PV}) \quad (16)$$

where $I_b$ is the battery current. The total power consumption given by Equation (16) can be expressed in terms of the electromechanical parameters as

$$P_{ci} = \left(\frac{R_a T_i}{k_t} + k_e \omega_{pi}\right)\left(\alpha \frac{T_i}{k_t} - I_{PV}\right) \quad (17)$$

Here, $\alpha = 1$ corresponds to the Drive Mode, and $\alpha = -1$ for the Regenerative Mode of the motor. If $\alpha$ is set to the fixed value -1, the regenerative motor functions only as a generator. The regenerative case can be implemented for the diving mode of the vehicle or a threaded/supported vehicle where the electric motors can operated as generators and charge onboard batteries. The negative power value in Equations (16) and (17) indicates the generation mode, and the positive is for power consumption.

There are many techniques to control electric motors. A typical approach is the use of Electronics Speed Controllers (ESC) which accept Pulse Width Modulation (PWM) signals from an onboard flight controller (microcontroller) to drive the motor. PMW signals provide a percentage of the battery voltage duty cycle by switching the signal sent to the motor on and off. The on-off pattern simulates the voltage between the peak voltage and zero proportional to the on portion of the signal versus off portion. Therefore, the power for each motor can be obtained by

$$P_{ci} = \%Duty\ Cycle \times v_a I_{ai}$$

where the percent duty cycle is the percentage of the time on versus off, sent from the ESCs to the motors.

The output current of the PV cells for the equivalent PV circuit in Figure 5 can be given by

$$I_{PV} = I_{SC} - I_d - I_{SH} \quad (18)$$

where $I_{PV}$ denotes the current applied to the load in the PV circuit, $I_{SC}$ is the short circuit current, $I_d$ is the diode current (leakage currents arise), and $I_{SH}$ is the current through the shunt resistor, expressed by the following equations

$$I_d = I_0(\exp(q_e(v_a + I_{PV}R_s)/n_i k_B T_c) - 1) \quad (19)$$

$$I_{SH} = (v_a + I_{PV}R_s)/R_{SH} \quad (20)$$

where $R_{SH}$ is a parallel shunt resistance, $I_0$ is the reverse (dark) saturation current (the leakage current in the absence of light), $q_e$ is the electron charge (1.602×10$^{-19}$ C), and $v_a$ is the voltage across the diode. The voltage drop in the electrical contacts is modeled as series resistance $R_s$. $k_B$ is Boltzmann's constant (1.381×10$^{-23}$ J/K), $T_c$ is the junction temperature (K), and $n_i$ is the ideality factor with $n_i$=1 for an ideal diode. The ideality factor varies from 1 to 2 depending on the fabrication process and the semiconductor material. An example of current-voltage (I-V) and power-voltage (P-V) curves for direct Sun radiation to a 2 m$^2$ of monocrystalline semi flexible silicon solar panels with 22.5% efficiency for the PV cells is given in Figure 6 (The specifications for the PV cells in the figure include size: 125 mm x125 mm; power output: 3.42 Watts, for 5.93 A current and 0.58 V voltage; Mass: 7 g).

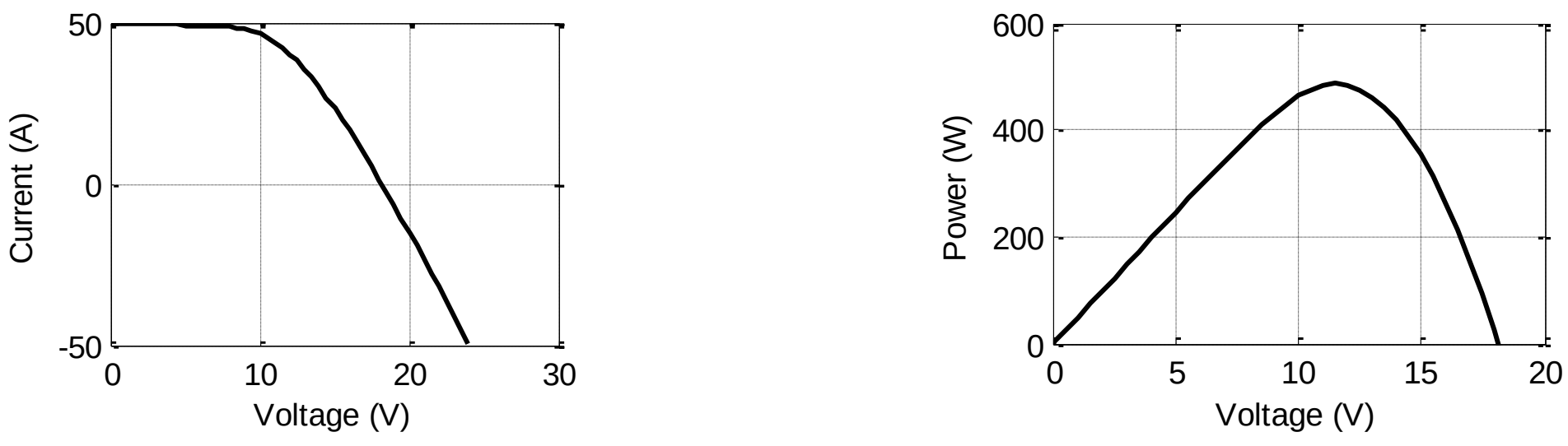

Figure 6. I-V and P-V curves of the PV cell.

The approximate available solar irradiance on a horizontal surface at ground level in direct Sun exposure with zero zenith angle in summer can be approximated as 1 kW/m$^2$. The current PV technology can offer efficiency of 5% to just above 40% depending on the type of PV cells. Flexible PV cells are normally suitable for installation on (aerial) vehicles' surfaces. The efficiency of flexible PV cells usually range from 5% to below 30%. Therefore typical available solar power for a vehicle range from 5% to 25% of 1 kW/m$^2$ with the current technology. The study in this paper is based on 1% to 40% efficiency. An overall efficiency, $\eta$, for the PV system is considered here that includes the available irradiance (including the angle of incidence, temperature, altitude, and time of the day and year), PV cells efficiency, and the electrical power system efficiency. Therefore, the lower bound of 1% efficiency is considered here for taking into account the overall efficiency of the system and not the PV cells efficiency alone. The value of $\eta$ can be calculated by individual efficiency of each element or component as $\eta_1 \times \eta_1 \times \cdots \times \eta_n$ (if $n$ components are considered). Due to many sources of uncertainties in the input and parameters of the system, Optimal Uncertainty Quantification technique, explained in the next section, is used which provides an optimal solution while taking into account incomplete and uncertain information in analyzing the behavior of the vehicle.

## V. Optimal Uncertainty Quantification

The condition of self-powering is achieved in a system when the available power from renewable or regenerative sources is equal to or larger than the power consumption of the system. A nondimensional power term $P_{non}$ can be realized in evaluating the self-powered condition. The nondimensional power is the UAV flight power demand divided by the generated power from solar energy. The problem of self-powered UAV system is formulated in a framework of Optimal Uncertainty Quantification (OUQ) [48] as a well-defined optimization corresponding to extremizing the probabilities of system failure with respect to the energy supply subject to the imposed constraints. OUQ takes into account uncertainty measures with optimal bounds and incomplete information in the energy inputs and parameters of the response function. The input energy and excitations, and output associated with a self-powered vehicle can include $I_{PV}$, $\boldsymbol{r}$, $\boldsymbol{v_B}$, $\dot{\boldsymbol{v}}_{\boldsymbol{B}}$, $\boldsymbol{\omega}_B$, $\dot{\boldsymbol{\omega}}_B$, $\mathbf{F}_B$, $\mathbf{M}_B$ which are generally constrained and uncertain. The input energy, for instance by considering $I_{PV}$, is required to provide the power for the actuation forces and moments, $\mathbf{F}_B$ and $\mathbf{M}_B$, through the controller parameters (e.g., PID gains) to produce a set of desired performances in terms of $\boldsymbol{r}$, $\boldsymbol{v_B}$, $\dot{\boldsymbol{v}}_{\boldsymbol{B}}$, $\boldsymbol{\omega}_B$, and $\dot{\boldsymbol{\omega}}_B$. There exist bounded ranges of the input parameters to the system, while the output is also constrained according to the desired performance of the system. The self-powered condition is analyzed in the framework of OUQ as discussed below. To formulate this problem, $P_{non}$, is the output. $I_{PV}$, and $\boldsymbol{r}$ are the inputs while the reference to fulfil the desired performance of the system can be evaluated by $\boldsymbol{r}$, $\boldsymbol{v_B}$, $\dot{\boldsymbol{v}}_{\boldsymbol{B}}$, $\boldsymbol{\omega}_B$, $\dot{\boldsymbol{\omega}}_B$. The required performance is achieved by the system's actuator forces and moments, $\mathbf{F_B}$ and $\mathbf{M}_B$, thorough the controller parameters (e.g., PID gains that produce the actuation forces and moments). The inputs can also include the parameters of the UAV system, and the external excitations such as wind load. The UAV control system which illustrates the relation between the parameters is shown in Figure 7.

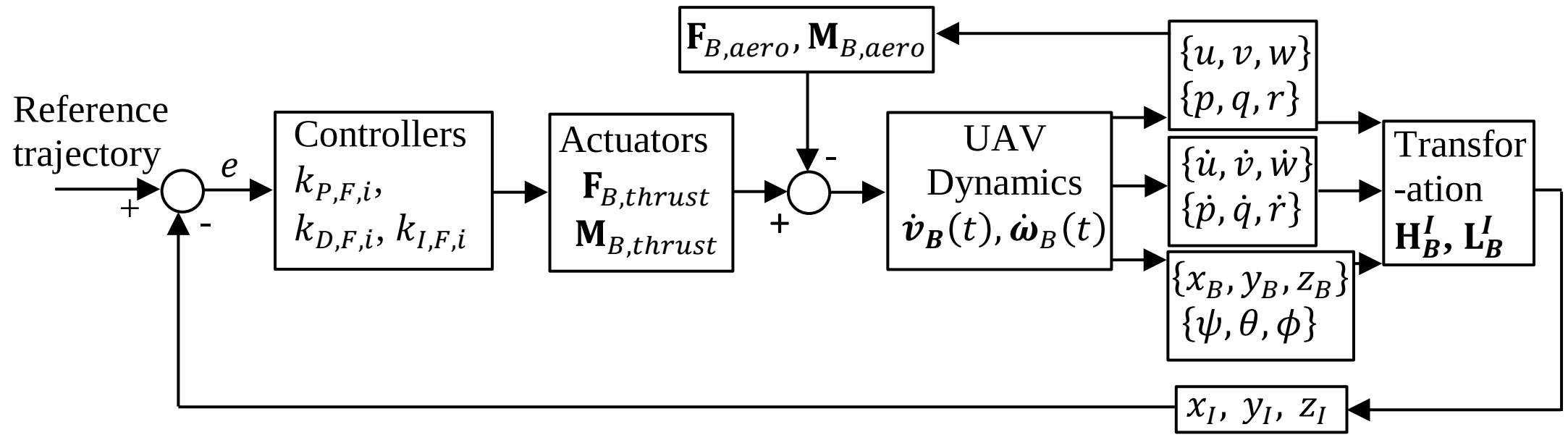

Figure 7. Feedback control diagram of the UAV.

The condition of self-powered flight is attained when the nondimensional power $P_{non}$, as given by the power consumption of the UAV $P_c$ divided the generated PV power $P_g$, is equal to or less than one, or $P_{non} = P_c/P_g \leq 1$. Assume $P_{non}: \mathcal{X} \to \mathbb{R}$, $X \to P_{non}(X)$, with the probability of $\mathbb{P} \in \mathcal{M}(\mathcal{X})$, where $X$ and $\mathcal{X}$ denote the deterministic and the stochastic ranges of the inputs, respectively. In a stochastic representation of a self-powered system, it is required that the probability of the nondimensional power transfer function, $P_{non}(X)$ becoming greater than one, which corresponds to failure of the self-powered scheme, to be less than $\epsilon$. This approach can be formulated as

$$\mathbb{P}[P_{non}(X) > 1] \leq \epsilon \tag{21}$$

Assume that the probability function is a member of admissible extremal scenarios $\mathcal{A}$ (or $(P_{non}, \mathbb{P}) \in \mathcal{A}$), where $\mathcal{A}$ is defined as

$$\mathcal{A} := \left\{ (p_{non}, \mu) \middle| \begin{array}{c} p_{non}: \mathcal{X}_1 \times \cdots \times \mathcal{X}_m \to \mathbb{R} \\ \mu = \mu_1 \otimes \mu_2 \otimes \cdots \otimes \mu_m \\ \mathbb{E}_\mu[P_{non}] \leq 1 \end{array} \right\} \tag{22}$$

$\mathcal{X}_j$ (or $\boldsymbol{\mathcal{X}}_j$ as a vector) denotes the inputs for $m$ inputs, $j = 1, \cdots, m$. $\mathbb{E}_\mu[P_{non}]$ is the bounded mean output power ($P_{non}$ can be used instead, if mean value is not needed), $\mu_j$ is the probability measure of the input parameter $\mathcal{X}_j$ ($\mu_j \in \mathcal{P}(\mathcal{X}_j)$), and $p_{non}$ is a possible output function of $P_{non}$ for the corresponding inputs/parameters $\mathcal{X}_j$. The original problem entails optimizing over a collection of $(p_{non}, \mu)$ that could be $(P_{non}, \mathbb{P})$. If $P_{non,max}$ and $P_{non,min}$ defined as the output upper bound and lower bound of $P_{non}$, is a function of upper and lower bounds of the consumed power and PV power, if known, the optimization problem is described as follows.

The optimal bounds on the probability of nondimensional power by the upper bound, $\mathcal{U}(\mathcal{A})$ (or $P_{non,max}$) can be given by

$$\mathcal{U}(\mathcal{A}) := \sup_{(p_{non},\mu)\in\mathcal{A}} \mu[P_{non}(X) > 1\ ]$$

and the lower bound, $\mathcal{L}(\mathcal{A})$ (or $P_{non,min}$), corresponding to the minimum required power consumption can be stated as

$$\mathcal{L}(\mathcal{A}) := \inf_{(p_{non},\mu)\in\mathcal{A}} \mu[P_{non}(X) > 1]$$

which give the optimal bounds as

$$\mathcal{L}(\mathcal{A}) \leq \mathbb{P}[P_{non}(X) > 1] \leq \mathcal{U}(\mathcal{A}) \tag{23}$$

The nondimensional power can be obtained by solving the constrained optimization problem for extremal scenarios $\mathcal{A}$, with the inputs/parameters also constrained. The inputs, $\mathcal{X}_j$ and $\mu_j$, can be constrained values with corresponding lower and upper bounds for each $j$. The input variables are indicated by $\mathcal{X}$, and the input vectors by $\boldsymbol{\mathcal{X}}$. The upper and lower bounds of the system inputs can be considered as

$$I_{PV} \in \mathcal{X}_1 := [I_{min}, I_{max}]\ \text{Amp},$$

In addition to the solar power as an input, the feedback control desired references, in terms of dynamic behavior including trajectory-tracking and path-following problems, can be evaluated as reference parameters, with the given bounds as

$$\boldsymbol{r} \in \boldsymbol{\mathcal{X}}_2 := [\boldsymbol{r}_{min}, \boldsymbol{r}_{max}]\ \text{m},$$

$$\boldsymbol{v_B} \in \boldsymbol{\mathcal{X}}_3 := \left[\boldsymbol{v_B}_{min}, \boldsymbol{v_B}_{max}\right]\ \text{m}/s,$$

$$\dot{\boldsymbol{v}}_{\boldsymbol{B}} \in \boldsymbol{\mathcal{X}}_4 := \left[\dot{\boldsymbol{v}}_{\boldsymbol{B}min}, \dot{\boldsymbol{v}}_{\boldsymbol{B}max}\right]\ \text{m/s}^2,$$

$$\boldsymbol{\omega}_B \in \boldsymbol{\mathcal{X}}_5 := \left[\boldsymbol{\omega}_{B_{min}}, \boldsymbol{\omega}_{B_{max}}\right]\ \text{rad/s},$$

$$\dot{\boldsymbol{\omega}}_B \in \boldsymbol{\mathcal{X}}_6 := \left[\dot{\boldsymbol{\omega}}_{B_{min}}, \dot{\boldsymbol{\omega}}_{B_{max}}\right]\ \text{rad/s}^2,$$

The above parameters can be formulated as part of the problem, if required. The disturbance to the system can be the aerodynamic forces as

$$\mathbf{F}_{B,aero} \in \boldsymbol{\mathcal{X}}_7 := \left[\mathbf{F}_{B,aero_{min}}, \mathbf{F}_{B,aero_{max}}\right]\ \text{N},$$

$$\mathbf{M}_{B,aero} \in \boldsymbol{\mathcal{X}}_8 := \left[\mathbf{M}_{B,aero_{min}}, \mathbf{M}_{B,aero_{max}}\right]\ \text{N.m},$$

The desired output associated with the reference inputs (e.g. $\boldsymbol{r}$), and energy input, $P_g$, can be achieved by the actuation forces and moments given by

$$\mathbf{F}_{B,thrust} \in \boldsymbol{\mathcal{X}}_9 := \left[\mathbf{F}_{B,thrust_{min}}, \mathbf{F}_{B,thrust_{max}}\right]\ \text{N},$$

$$\mathbf{M}_{B,thrust} \in \boldsymbol{\mathcal{X}}_{10} := \left[\mathbf{M}_{B,thrust_{min}}, \mathbf{M}_{B,thrust_{max}}\right]\ N.m,$$

In analyzing the problem, $\mathbf{F}_{B,thrust}$ and $\mathbf{M}_{B,thrust}$ can be considered as the actuation forces directly, or alternatively the proportional, derivative, and integral feedback control gains that generates these forces and moments can be used in the optimization problem. Therefore, the upper and lower limits of the actuation forces can be assessed as functions of control parameters by

$$k_{P,F,i} \in \mathcal{X}_{P,F,i} := \left[k_{P,F,i,min}, k_{P,F,i,max}\right], k_{D,F,i} \in \mathcal{X}_{D,F,i} := \left[k_{D,F,i,min}, k_{D,F,i,max}\right], \quad k_{I,F,i} \in \mathcal{X}_{I,F,i} := \left[k_{I,F,i,min}, k_{I,F,i,max}\right] \tag{24}$$

where $k_{P,F,i}$ is the proportional control gain associated with the propeller thrust force, $f_i$ (Figure 4), of motor $i$. $\mathcal{X}_{P,F,i}$ is the indication that the nondimensional power is a function of the stochastic form of $k_{P,F,i}$ in the optimization process. Similarly, $k_{D,F,i}$ and $k_{I,F,i}$ are the derivative and integral control gains, respectively, associated with propeller thrust force of motor $i$, and $\mathcal{X}_{D,F,i}$ and $\mathcal{X}_{I,F,i}$ are the corresponding stochastic inputs to the optimization problem.

$$k_{P,\boldsymbol{\alpha},i} \in \mathcal{X}_{P,\boldsymbol{\alpha},i} := \left[k_{P,\boldsymbol{\alpha},i,min}, k_{P,\boldsymbol{\alpha},i,max}\right], k_{D,\boldsymbol{\alpha},i} \in \mathcal{X}_{D,\boldsymbol{\alpha},i} := \left[k_{D,\boldsymbol{\alpha},i,min}, k_{D,\boldsymbol{\alpha},i,max}\right], \quad k_{I,\boldsymbol{\alpha},i} \in \mathcal{X}_{I,\boldsymbol{\alpha},i} := \left[k_{I,\boldsymbol{\alpha},i,min}, k_{I,\boldsymbol{\alpha},i,max}\right] \tag{25}$$

where $k_{P,\boldsymbol{\alpha},i}$ is the proportional control gain associated with the rate of tilt angle $\boldsymbol{\alpha}$ (Figure 4) of propeller $i$. $\mathcal{X}_{P,\boldsymbol{\alpha},i}$ is the indication that the nondimensional power is a function of $k_{P,\boldsymbol{\alpha},i}$ in the optimization process. Similarly, $k_{D,\boldsymbol{\alpha},i}$ and $k_{I,\boldsymbol{\alpha},i}$ are the derivative and integral control gains, respectively, associated with rate of tilt angle of propeller $i$, and $\mathcal{X}_{D,\boldsymbol{\alpha},i}$ and $\mathcal{X}_{I,\boldsymbol{\alpha},i}$ are the corresponding inputs to the optimization problem.

Therefore, the admissible set $\mathcal{A}$ may be given by

$$\mathcal{A} := \left\{ (p_{non}, \mu) \middle| \begin{array}{c} p_{non}: \mathcal{X}_1 \times \boldsymbol{\mathcal{X}}_2 \times \cdots \times \boldsymbol{\mathcal{X}}_{10} \to \mathbb{R} \\ \mu = \mu_1 \otimes \mu_2 \otimes \cdots \otimes \mu_{10} \\ \mathbb{E}_\mu[P_{non}] \leq 1 \end{array} \right\} \tag{26}$$

or if the problem is defined on the basis of controllers gains, $\mathcal{A}$ can be expressed as

$$\mathcal{A} := \left\{ (p_{non}, \mu) \middle| \begin{array}{c} p_{non}: \mathcal{X}_1 \times \mathcal{X}_{P,F,i} \times \mathcal{X}_{D,F,i} \times \mathcal{X}_{I,F,i} \times \mathcal{X}_{P,\boldsymbol{\alpha},i} \times \mathcal{X}_{D,\boldsymbol{\alpha},i} \times \mathcal{X}_{I,\boldsymbol{\alpha},i} \to \mathbb{R} \\ \mu = \mu_1 \otimes \mu_{1,i} \otimes \mu_{2,i} \otimes \mu_{3,i} \otimes \mu_{4,i} \otimes \mu_{5,i} \otimes \mu_{6,i} \\ \mathbb{E}_{\mu}[P_{non}] \le 1 \end{array} \right\} \quad (27)$$

where $\mu_1$ is the probability measure of the input parameter $\mathcal{X}_1$ ($\mu_1 \in \mathcal{P}(\mathcal{X}_1)$), and $\mu_{1,i}$ is the probability measure of the control gains for propeller/motor $i$.

The optimization output cost function can include multiple parameters. For instance, it can consist of the displacement associated with trajectory-tracking and path-following problems in addition to the nondimensional power.

If (experimental) sample data is available then the OUQ can be extended to Machine Wald [49] technique which is equivalent to performing Bayesian inference but optimizing the prior. In Machine Wald, if an estimation of a function $\Phi(\mu)$, is function $\theta$ of sample data $d$, then estimation of error $\theta(d) - \Phi(\mu)$ is required to tend to zero. A game theoretic approach may be applied when Player I provides known input, output and system parameter information, and Player II provides some uncertain information about the system [50]. In this problem, an optimal solution in regards with the performance of the system and achieving self-powered condition is required.

Examples of developing fully self-powered dynamic systems associated with solar-powered UAVs in the framework of OUQ are discussed in the following sections.

## VI. Solar-powered speed

Solar-powered speed is defined here as the flight speed that the vehicle can sustain when the source of power is solar energy, as received from the photovoltaic (PV) cells in real time, for a neutrally buoyant vehicle. In this section, the solar-powered speed is investigated for vehicles with various geometries, including cuboid and ellipsoid geometries. In this problem, the condition achieving for solar powered speed is $P_{non} = P_c/P_g \le 1$. For a one-dimensional problem with constant wind relative speed, the power $P_c$ that is required to propel a vehicle at a constant speed $v_{air}$ can be given by Equations (1) and (6), and consequently the power is $P_c = \frac{1}{2}\rho C_d A v_{air}^3$. In this equation, $\rho$ denotes the air density, and $C_d$ is the drag coefficient of the geometry of the vehicle. The source of energy for generating this power is solar ($P_g$). In this one-dimensional problem the wind and thrust are collinear. As stated in the OUQ section, the parameters of the system can be introduced as uncertain inputs to a problem. The drag coefficient is highly dependent on the geometry of the vehicle, and Reynolds number. In order to obtain a feasible solar-powered speed an upper bound value of $C_d$, $C_{d_{max}}$, is considered corresponding to a Reynolds number of $10^4 \le R_e \le 10^7$. This upper bound $C_d$ also takes into account the drag due to skin friction as part of the form drag. Therefore, to express this problem in the OUQ context we have

$$v_{air} \in \mathcal{X}_1 := [v_{min}, v_{max}] \text{ m}/s,$$

$$C_d \in \mathcal{X}_2 := [C_{d_{min}}, C_{d_{max}}] \text{ m}/s,$$

$$\mathbb{P}[P_{non}(X) > 1] \le 0$$

Thus, the admissible set is

$$\mathcal{A} := \left\{ (p_{non}, \mu) \middle| \begin{array}{c} p_{non}: \mathcal{X}_1 \times \mathcal{X}_2 \to \mathbb{R} \\ \mu = \mu_1 \otimes \mu_2 \\ P_{non} \le 1 \end{array} \right\}$$

Zero angle of attack of the vehicle is assumed in this model. However, the approach here can be extended to $C_d$ values with any angle of attack. Additional parameters can be considered as uncertain and bounded values, if needed.

Note that the worst case (the speed associated with maximum drag) is considered for the uncertainty bounds (i.e., $C_{d_{max}}$) in this problem. As a one-dimensional problem with constant velocity, the numerical solution for $v$ is found from $\frac{P_c}{P_g} = 1$, where $P_g$ is discussed below.

The power required for achieving solar-powered speed is supplied by solar energy. This speed is investigated for cuboid and ellipsoid shapes as baseline designs associated with the buoyancy hull envelope of aerial vehicles. There is no induced drag as the shapes under consideration are symmetric (i.e., neutral camber) where no lift is generated. It is assumed that the required lift force for maintaining the altitude is provided by buoyancy of the vehicles' hull filled with lighter than air gas. Constant-altitude and constant velocity cruise flight is considered here.

Cuboid: Brunel Solar Quadrotor UAV (Figure 1) with dimensions $a \times b \times L$, $a$ = 2 m, $b$ = 1 m, $L$ = 3 m, is an example of a cuboid geometry. PV cells are placed on the top surface with the (maximum) dimensions of $a \times L$. The solar-powered (wind relative) speed (airspeed), which is denoted by $v_{solar}$ here, is obtained by solving for $v_{air}$ in $P_{non} = 1$ (i.e., $P_c = P_g$) as

$$v_{solar} = \left[\eta I_{PV} V_{PV} / \left(0.5 \rho C_{d_{max}} ab\right)\right]^{1/3} \quad (28)$$

where $\eta$ denotes the overall PV system efficiency, and $V_{PV}$ is the PV cells voltage. The power supplied by the PV cells, $P_g = I_{PV} V_{PV}$, can be given as a factor of 1 kW/m$^2$, multiplied by the PV cell surface area, and by taking into account an overall efficiency of $\eta$ (this efficiency can be considered as a bounded uncertain input for achieving more accuracy). Therefore, the PV power, when the top surface of the cuboid is covered with the cells, is given by $1000\eta aL$. Equation (28) with upper bound of drag coefficient of $C_{d_{max}}$= 2 can be expressed as $v_{solar} = [1000\eta L/(\rho b)]^{1/3}$. The solar-powered (wind relative) speed (or airspeed) with respect to $L/b$, for various $\eta$ in percentage is plotted in Figure 8.

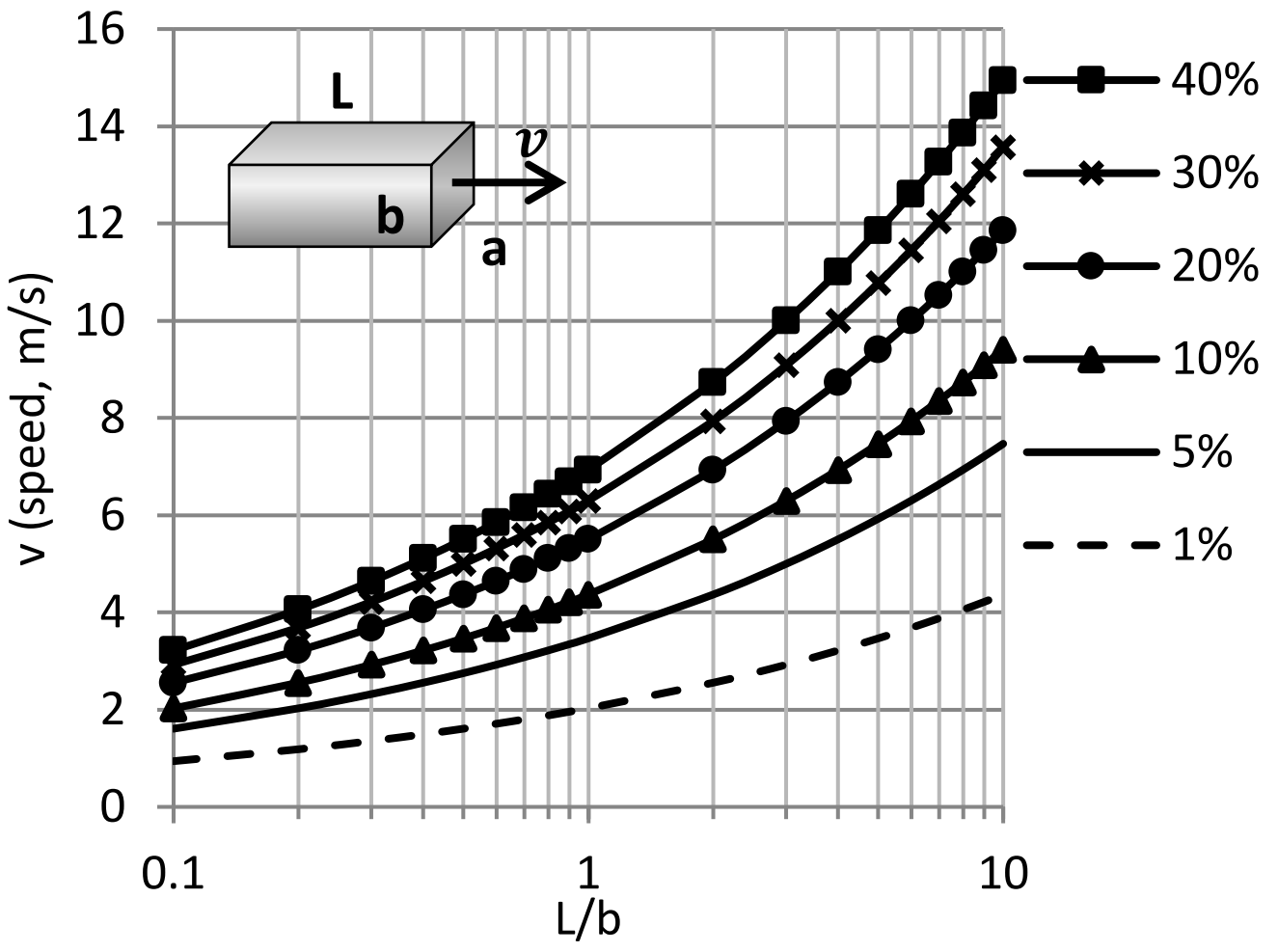

Figure 8. Solar-powered speed for cuboid versus $\boldsymbol{L/b}$ for various $\boldsymbol{\eta}$ values.

For a constant volume of $a \times b \times L$, preferred dimensions that are aerodynamically efficient and provide larger surface areas for the PV cells are based on $b \leq a \leq L$.

Based on the dimensions of the Brunel Solar Quadrotor UAV, $L/b$ = 3, and by referring to Figure 8, if the entire top surface area of $a \times L$ = 6 m$^2$ is covered with PV cells with approximate overall PV system efficiency of $\eta$ = 20%, and exposed to direct solar irradiance, the solar-powered speed will be $v$ = 7.9 m/s. Note that the surfaces for Brunel Solar Quadrotor UAV is slightly curved which gives smaller drag force. This has not been considered in the simulation. Also, as the top surface is not fully covered with PV cells, the total PV cell area needs to be updated. Therefore, the solar-powered speed ($v_{solar} = \left[1000\eta A_{PV}/\left(0.5\rho C_{d_{max}}(C_{d_{new}}/C_d)ab\right)\right]^{1/3}$) is obtained using Figure 8 by applying an updated $L/b$ value, for Brunel UAV, given by $(L/b)_{updated}$ = $2A_{PV}/(C_{d_new}ab)$, where $A_{PV}$ is the total PV cells area, $C_{d_new}$ is the drag coefficient of the curved geometry (approximately equal to 1), and number 2 in the numerator is the estimated upper bound drag coefficient that was originally used, $C_{d_{max}}$. 30 PV cells with each cell covering the area of 0.125 m × 0.125 m is used for the Brunel UAV which gives 0.47 m$^2$ PV surface area with $(L/b)_{updated}$ = 0.47. This updated $L/b$ brings the solar-powered speed to about 4 m/s. This procedure shows how Figure 8 can be used as a generalized reference plot when the geometry is not completely cuboid, and the PV cell area, $A_{PV}$, is smaller than the total top surface of the cuboid.

Ellipsoid: The procedure discussed above is repeated here for ellipsoid shapes. An example of an ellipsoid hull (or spheroid in this case) is the Brunel Solar Octorotor UAV in Figure 2, with dimensions $D$ = 1.6 m, $b$ = 2.5 m, $L$ = 2.5 m, where $b$ is the width of the body. PV cells are placed on the top curved surface area. The solar irradiance of PV cells is reduced due to the curvature of the surface and the increase of angle of incidence. A projected area of $\pi bL/4$ is considered for the PV surface area as an approximation. This approximation can be taken into account in the overall PV system efficiency for more detailed analysis. The frontal area is $\pi bD/4$. The solar-powered (wind relative) speed (airspeed) can be obtained by

$$v_{solar} = \left[\eta I_{PV} V_{PV}/\left(0.5\rho C_{d_{max}} \pi bD/4\right)\right]^{1/3} \tag{29}$$

The PV power, when the top surface of the ellipsoid is covered with the cells and exposed to direct solar irradiance, can be estimated by $1000\eta\pi bL/4$. Equation (29) with assuming the upper bound drag coefficient of $C_{d_{max}}$= 1 can be expressed as $v_{solar} = [1000\eta L/(0.5\rho D)]^{1/3}$. This solar-powered speed with respect to $L/D$, for various $\eta$ in percentage is plotted in Figure 9. The range $L/D < 1$ may not be practical. However, it is plotted for cases when the $(L/D)_{updated}$ is used as discussed below. Note that these updated $L/D$ values or $(L/D)_{updated}$ do not carry the same physical representation of $L/D$, and only used for generalizing the plot as a reference as explained below.

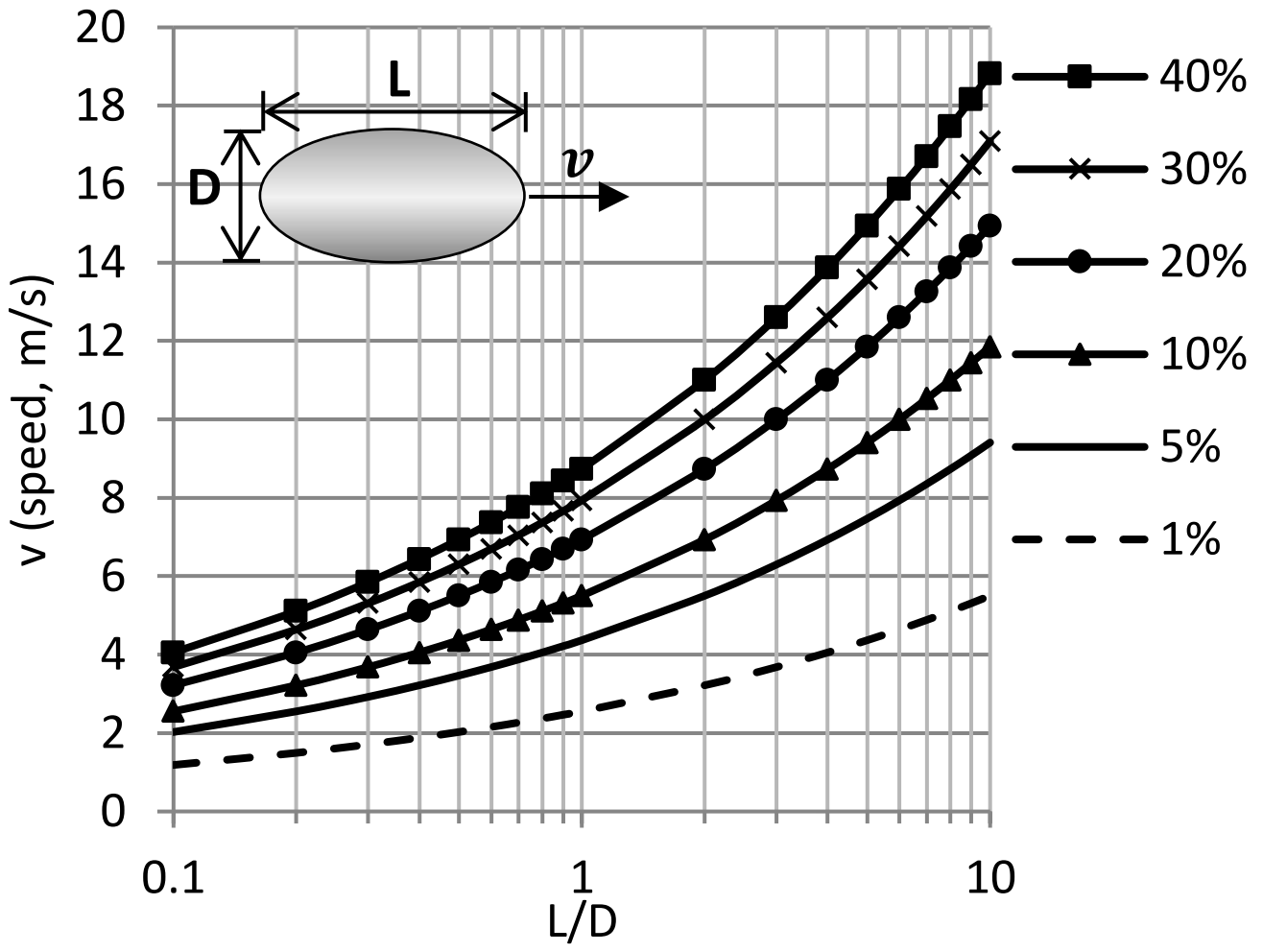

Figure 9. Solar powered speed for ellipsoid versus $\boldsymbol{L/D}$ for various $\boldsymbol{\eta}$ values.

For a constant volume of $4/24 \times \pi \times b \times D \times L$, preferred dimensions that are aerodynamically efficient and provide larger surface areas for the PV cells are based on $D \leq b \leq L$.

Based on the dimensions of the Brunel Solar Octorotor UAV, $L/D$=1.5625, if the entire projected top surface area of $\pi bL/4$ = 4.90 m$^2$ is covered with PV cells with an approximate overall PV system efficiency of $\eta$ = 5%, and exposed to direct solar irradiance, the solar-powered speed will be $v$ = 5.07 m/s, and by referring to Figure 9. Note that the top surface is not fully covered with PV cells in this case and hence the total PV cell area needs to be updated. To update the values for partially covered PV surface area, The solar-powered speed ($v_{solar} = [1000\eta A_{PV}/(0.5\rho C_d \pi bD/4)]^{1/3}$) is obtained using Figure 9 by applying an updated $L/D$ value, for the Brunel UAV, given by $(L/D)_{updated}$ = $A_{PV}/(\pi bD/4)$, where $A_{PV}$ is the total PV cells area. 26 PV cells with each cell having the area of 0.076 m × 0.305 m is used for the Brunel UAV which gives 0.60 $m^2$ PV surface area with $(L/D)_{updated}$ = 0.19. This updated $L/b$ brings the solar-powered speed to about 2.5 m/s. Note that despite the fact that $(L/D)_{updated}$ does not serve as the same physical representation of $L/D$ but it allows Figure 9 to be used as a generalized reference for ellipsoid shape hulls with any PV cell area, $A_{PV}$.

If the upper bound drag coefficient of the Brunel Trirotor UAV (Figure 3) is assumed as $C_{d_{max}}$= 0.5, and the generalized ellipsoid approach presented above is followed, with dimensions $D$ = 1.10 m, $b_{max}$ = 1.750 m, $L$ = 2.10 m, and PV efficiency of $\eta$ = 8.9% exposed to direct solar irradiance, the solar-powered speed can be estimated as $v_{solar} = \left[1000\eta A_{PV}/\left(0.5\rho C_{d_{max}}(C_{d_{new}}/C_d)\pi bD/4\right)\right]^{1/3}$, with $(L/D)_{updated}$=$C_{d_{max}} A_{PV}/(C_{d_new} \pi bD/4)$ (here, $C_d$ = 1 is the original $C_{d_{max}}$ used in the plot of Figure 9, and $C_{d_new}$ = 0.5 for the Tri-rotor UAV), $A_{PV}$ = 0.432 $m^2$ for this UAV, which gives $(L/D)_{updated}$ = 0.57. This updated $(L/D)_{updated}$ gives the solar-powered speed of $v_{solar}$= 4.35 m/s, by interpolating for $\eta$ = 8.9% in the plot in Figure 9.

It should be added that the solar-powered speed values associated with Brunel UAVs can be improved by using larger PV cell areas and higher PV efficiencies. For large enough Reynolds numbers, $C_d$ can be as low as 0.005 which will lead to $(L/D)_{updated}$ = $C_d A_{PV}/(C_{d_new} \pi bD/4)$ = 57.1, and solar-powered speed of $v$ = 16.8197 m/s for $\eta$ = 5%, and $v_{solar}$ = 21.1915 m/s for $\eta$ = 10%, as in Figure 10. In this plot, the $L/D$ values are extended to 100. Although $L/D$ = 100 is not practical, it is used for updated cases of $(L/D)_{updated}$, which does not inherent the same physical representation of $L/D$ but it allows the calculation of solar-powered speed for various $C_d$, $A_{PV}$, and dimensions using the same plot as a generalized reference.

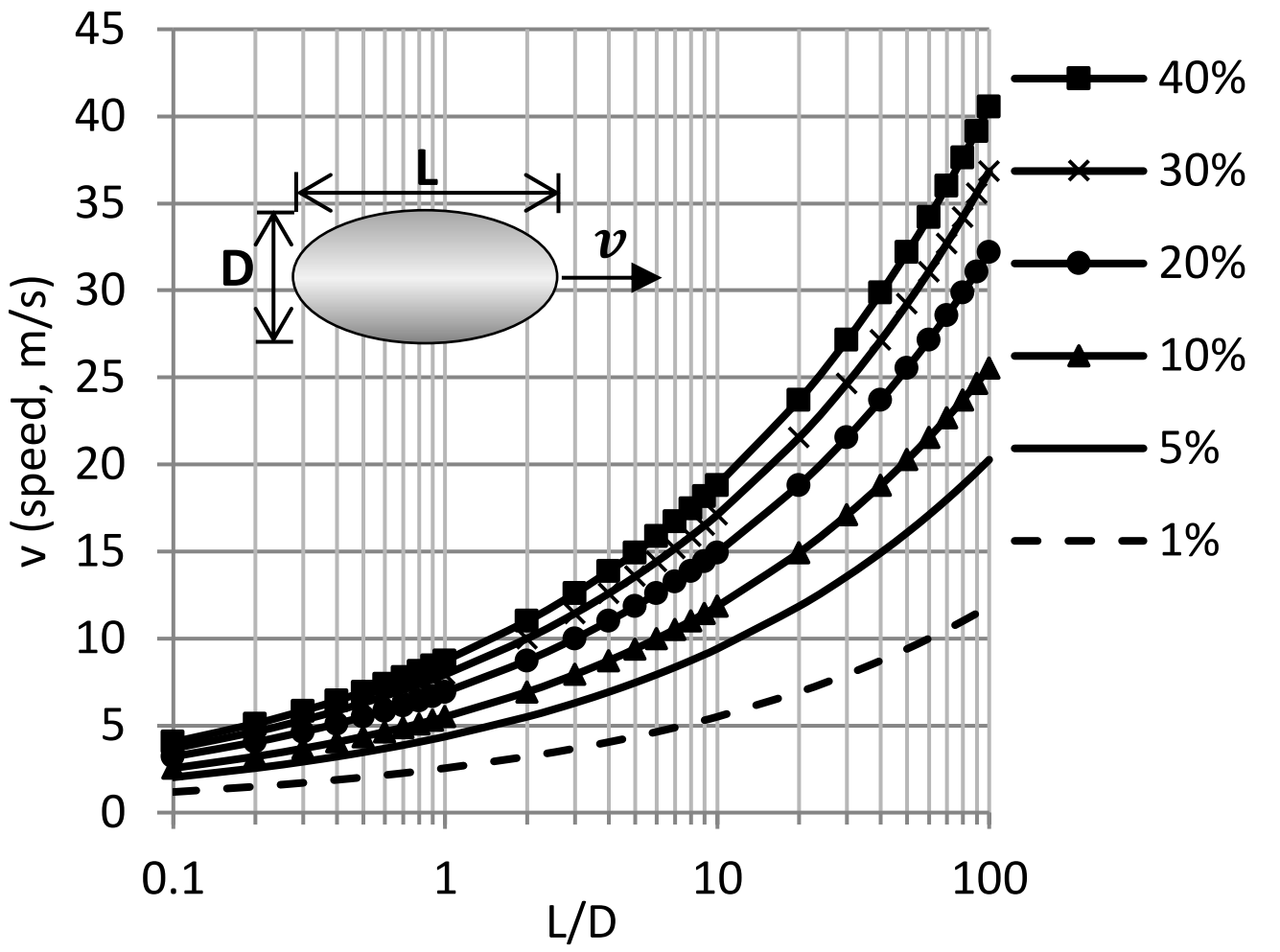

Figure 10. Solar-powered speed for ellipsoid versus $L/D$ for various $\eta$ values.

The above approach can be applied to airfoils, and other geometric shapes for various aerial vehicles.

The above procedure attempts to obtain solar-powered speeds as a function of PV cell characteristics, and the vehicles' size, shape, and drag force. In a design process, it is required to further address various aspects of the system as: a) sizing the motors, electronics and battery based on the power requirement; c) find the overall weight of the vehicles based on the motors, batteries, control units, electronics, the frame, and other related components; c) Calculate the volume required for buoyancy (based on the weight calculated above); d) If the size based on the calculated volume is not desirable then select larger or smaller size thorough an iterative process; e) If the size is smaller than desired, then calculate the extra weight and find the power required for lift (due to partial buoyancy); f) If the size is larger than required, then extra buoyancy force can be compensated by additional dead weight or reducing the size (area) for the PV cells.

It should be noted that the neutrally buoyant condition is assumed here in obtaining solar-powered speed (which in fact, it is normally the requirement for the LTA vehicles design). Therefore, if the vehicle is not neutrally buoyant, the associated power for the additional weight must be taken into account in the calculation of the power quantity.

The solar-powered speed procedure obtains a constant velocity powered by solar energy. In order to further extend the investigation, acceleration of the vehicle is taken into account. Equations (1) and (6) are employed again where the acceleration is non-zero. Therefore, $P_c = \frac{1}{2}\rho C_d A v_{air}^3 + m\dot{v}_{air} v_{air}$, and for $P_{non} = \frac{P_c}{P_g} = 1$, the upper bound of acceleration, or $a$ in Figure 11, versus the upper bound of $v_{air}$, for solar self-powered motion is given by Figure 11.

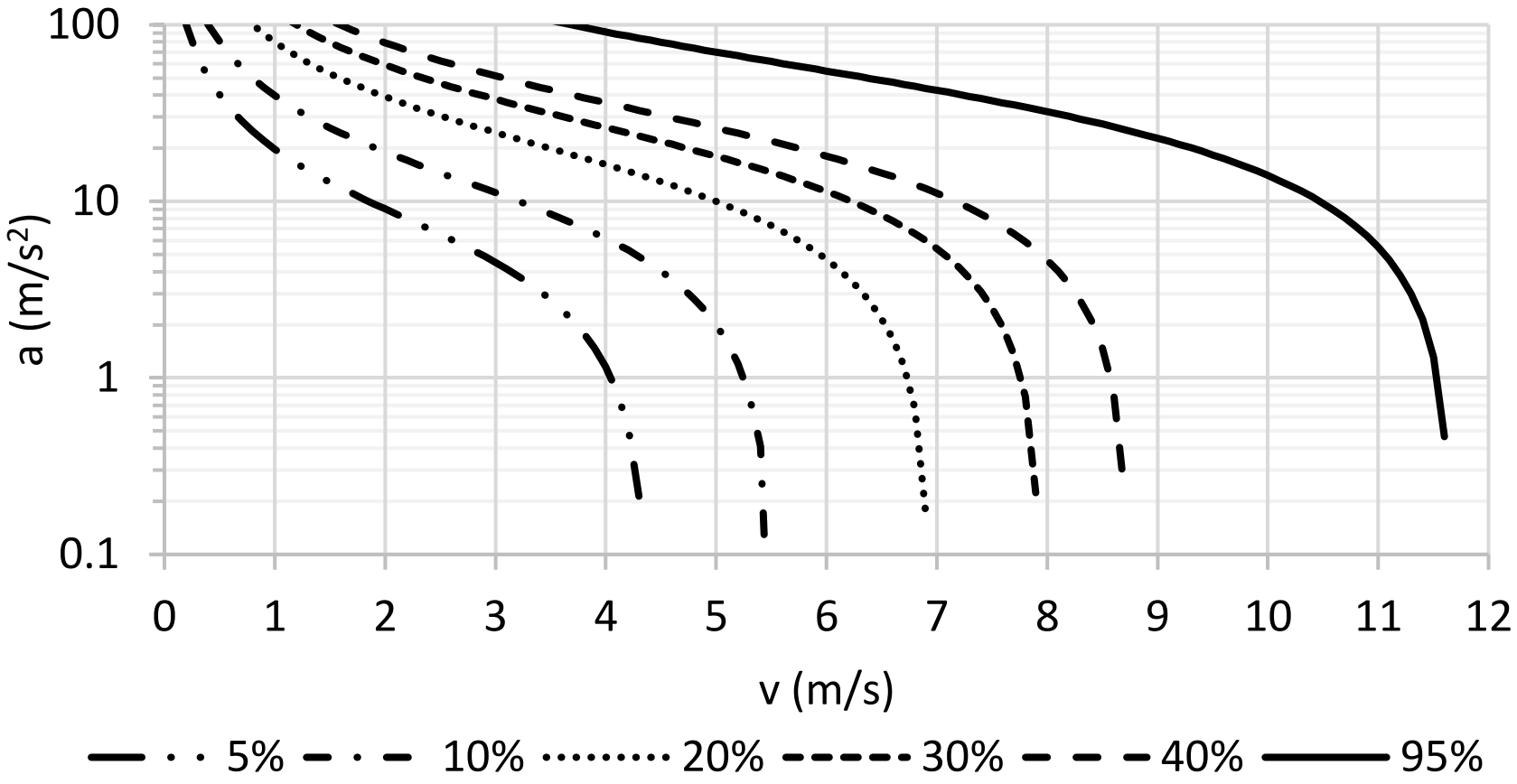

Figure 11. Upper bound accelerations and velocities for achieving solar self-powered condition.

The problem of solar-powered speed is extended to self-powered trajectory tracking in the next section.

## VII. Self-powered trajectory tracking

The self-powered control of vehicles for trajectory tracking while using solar energy in real time is explored in this section. This problem is formulated as below.

$$\begin{aligned} I_{PV} &\in \mathcal{X}_1 := [I_{min}, I_{max}] \text{ Amp}, \\ \boldsymbol{r} &\in \boldsymbol{\mathcal{X}}_2 := [\boldsymbol{r}_{min}, \boldsymbol{r}_{max}] \text{ m}, \\ \boldsymbol{v_B} &\in \boldsymbol{\mathcal{X}}_3 := [\boldsymbol{v_B}_{min}, \boldsymbol{v_B}_{max}] \text{ m}/s, \\ \dot{\boldsymbol{v}}_{\boldsymbol{B}} &\in \boldsymbol{\mathcal{X}}_4 := [\dot{\boldsymbol{v}}_{\boldsymbol{B}min}, \dot{\boldsymbol{v}}_{\boldsymbol{B}max}] \text{ m/s}^2, \\ \boldsymbol{\omega}_B &\in \boldsymbol{\mathcal{X}}_5 := [\boldsymbol{\omega}_{B_{min}}, \boldsymbol{\omega}_{B_{max}}] \text{ rad/s}, \\ \dot{\boldsymbol{\omega}}_B &\in \boldsymbol{\mathcal{X}}_6 := [\dot{\boldsymbol{\omega}}_{B_{min}}, \dot{\boldsymbol{\omega}}_{B_{max}}] \text{ rad/s}^2, \end{aligned} \tag{30}$$

where $\boldsymbol{r}$ is a candidate for the desired trajectory and $\boldsymbol{v_B}$, $\dot{\boldsymbol{v}}_{\boldsymbol{B}}$, $\boldsymbol{\omega_B}$, $\dot{\boldsymbol{\omega}}_{\boldsymbol{B}}$ can define the desired performance of the control system which is associated with overshoot, peak time, and settling time of the system response. Proportional, derivative, and integral feedback gains that control propulsion forces and moments are optimized to achieve self-powered trajectory tracking while considering the dynamic performance. The upper and lower limits of the actuation forces can be considered as functions of control parameters as

$$k_{P,F,i} \in \mathcal{X}_{P,F,i} := [k_{P,F,i,min}, k_{P,F,i,max}], \; k_{D,F,i} \in \mathcal{X}_{D,F,i} := [k_{D,F,i,min}, k_{D,F,i,max}],$$

$$k_{I,F,i} \in \mathcal{X}_{I,F,i} := [k_{I,F,i,min}, k_{I,F,i,max}],$$

Therefore, the admissible set $\mathcal{A}$ may be given by

$$\mathcal{A} := \left\{ (p_{non}, \mu) \middle| \begin{matrix} p_{non}: \mathcal{X}_1 \times \boldsymbol{\mathcal{X}}_2 \times \cdots \times \boldsymbol{\mathcal{X}}_6 \times \mathcal{X}_{P,F,i} \times \mathcal{X}_{D,F,i} \times \mathcal{X}_{I,F,i} \to \mathbb{R} \\ \mu = \mu_1 \otimes \cdots \otimes \mu_6 \otimes \mu_{1,i} \otimes \mu_{2,i} \otimes \mu_{3,i} \\ P_{non} \le 1 \end{matrix} \right\} \tag{31}$$

The upper and lower bounds represent the acceptable range of stochastic or incomplete information in regards with each of the parameters in Equation (31) and should result in the probability of $\mathbb{P}[P_{non}(X) > 1] \le \epsilon$ which is the state for the system not to fail in achieving self-powered condition. We also require $\epsilon$=0 in this problem which results in $P_{non} \le$1.

The longitudinal equations of motion using Equations (1) and (2) in two dimensions (vertical plane) are

$$\dot{u} = \frac{\mathbf{F}_B}{m} - g \sin\theta + \frac{B}{m} \sin\theta - qw$$

$$\dot{w} = \frac{\mathbf{F}_B}{m} + g \cos\theta - \frac{B}{m} \cos\theta + qu$$

$$\dot{q} = \mathbf{M}_B / \boldsymbol{I}_B$$

for translation and rotation, where $\dot{\theta} = q$, and $\dot{x}_I = u\cos\theta + w\sin\theta$, $\dot{z}_I = -u\sin\theta + w\cos\theta$. Assuming neutral buoyancy, for the case of Brunel Quadrotor UAV as an example in this section, we have $B = mg$, where $B$ is the buoyancy force. The feedback control system is as in Figure 7, where the propulsion given by the electric motors are $\mathbf{F}_{B,thrust} = k\ \omega_{p1}^2 + k\ \omega_{p2}^2$ using Equation (5), for $i$=1, 2 (i.e., two motors only). Accordingly, the moment is given by $\mathbf{M}_{B,thrust}= kl\ \omega_{p1}^2 \pm kl\ \omega_{p2}^2$, with positive or negative sign depending on the direction of the thrust vector. The properties of the UAV are considered as: and total mass of 11.3 kg including added mass, $\boldsymbol{I}_B$ = 2.76 kg.m$^2$, $l$ = 1.65 m, PV cells with 10% efficiency, and spherical hull with a radius of 1.25 m. The frontal area, and the total PV cell area are both calculated as $\pi(1.25)^2$.

The reference altitudes, $z_{ref}$, in this example are chosen as in Figure 12. Ramp reference inputs are considered for horizontal motion denoted by $x_{ref}$. Step and ramp inputs are often used in analysis of control systems. However, the choice of reference inputs in Figure 12 are merely used to demonstrate the self-powered condition. This is only for to demonstrate the self-powered condition with some examples. Of course, other trajectories can be used for analyzing the self-powered condition, if needed.

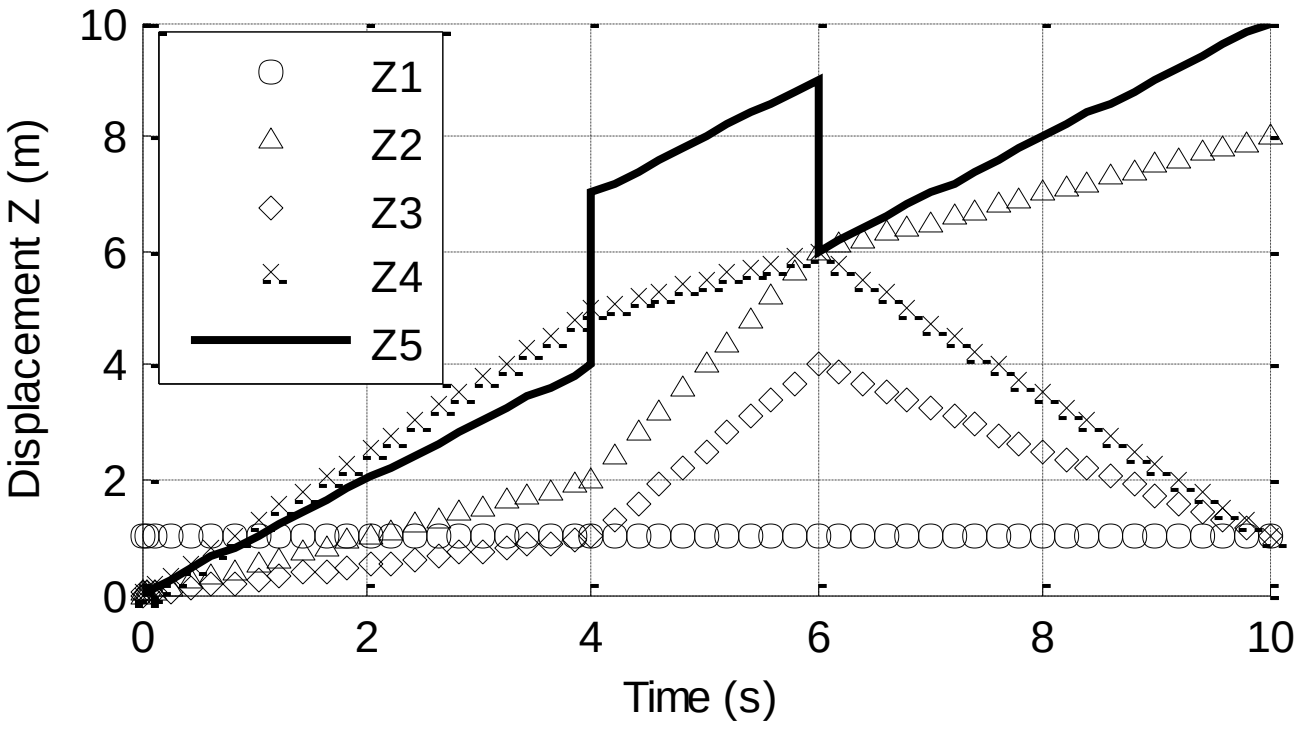

Figure 12. Reference altitudes $z_{ref}$.

The PID control gains that give $P_{non}$ ≤1 in real time for each duty cycle are given in Table 1. $x_I$, $z_I$, velocity, force, moment, and nondimensional power (denoted as "P" on the plots) are plotted for each $z_{ref}$, and corresponding optimized PID gains in Table 1. These plots are given in Figure 13 to Figure 18, where the velocity values are calculated by $(\dot{x}_B^2 + \dot{z}_B^2)^{1/2}$ in body frame, and actuators forces and moments are controlled by the PID gains given in Table 1.

Table 1. Controller gains associated with self-powered condition.

| $\boldsymbol{z_{ref}}$ | Step | | Ramp | | Z2 | | Z3 | | Z4 | | Z5 | |
|---|---|---|---|---|---|---|---|---|---|---|---|---|
| Force/Angle | F | $\theta$ | F | $\theta$ | F | $\theta$ | F | $\theta$ | F | $\theta$ | F | $\theta$ |
| P | 122.8 | 6.4 | 122.8 | 8 | 12.3 | 24.9 | 12.3 | 24.9 | 12.3 | 24.9 | 5.36 | 6.26 |
| I | 10.8 | 0.25 | 10.8 | 0.46 | 0.34 | 2.0 | 0.34 | 2.0 | 0.34 | 2.0 | 0.1 | 0.25 |
| D | 150.8 | 14.6 | 150.8 | 18.2 | 47.7 | 33.6 | 47.7 | 33.6 | 47.7 | 33.6 | 31.5 | 16.8 |
| Rise time (s) | 0.108 | 0.25 | 0.108 | 0.21 | 0.34 | 0.12 | 0.34 | 0.12 | 0.34 | 0.12 | 0.51 | 0.23 |
| Settling time (s) | 1.53 | 3.7 | 1.53 | 3.18 | 4.85 | 1.68 | 4.85 | 1.68 | 4.85 | 1.68 | 7.3 | 3.36 |
| Overshoot | 20% | 12% | 20% | 15% | 20% | 20% | 20% | 20% | 20% | 20% | 20% | 20% |

Figure 13 illustrates $x_I$, $z_I$, velocity, force, moment, and nondimensional power in response to reference inputs of $x_{ref}$ as a ramp function with a slope of 1, and $z_{ref}$ as a step (or Z1 in Figure 12). This procedure is repeated for various $x_I$ and $z_I$ in Figure 14 to Figure 18. In Figure 14, the reference inputs for $x_{ref}$ and $z_{ref}$ are both ramps with slopes of 1.

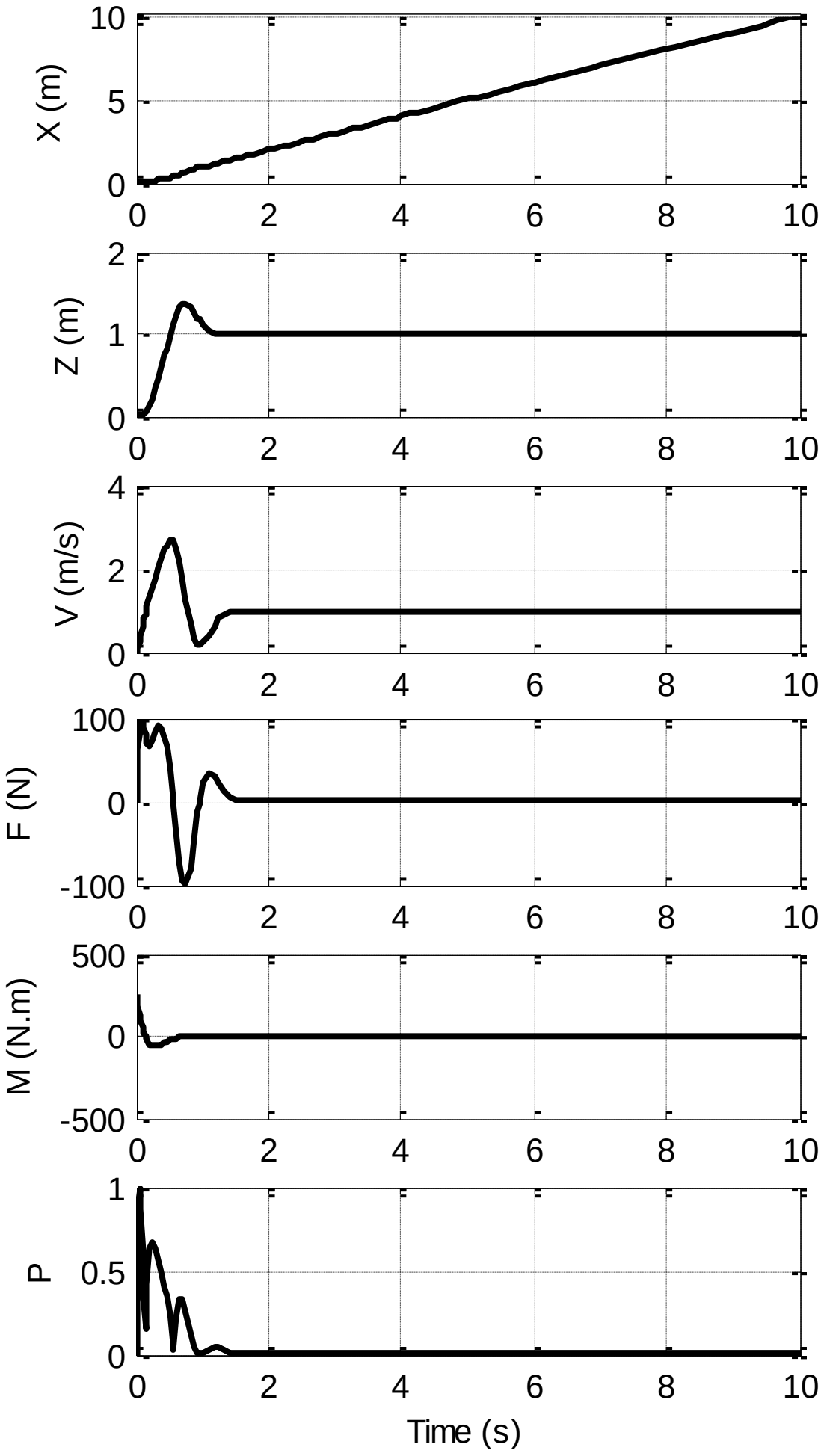

Figure 13. $x_I$, $z_I$, velocity, force, moment, and nondimensional power in response to reference inputs of $x_{ref}$ as a ramp function with a slope of 1, and $z_{ref}$ as a step.

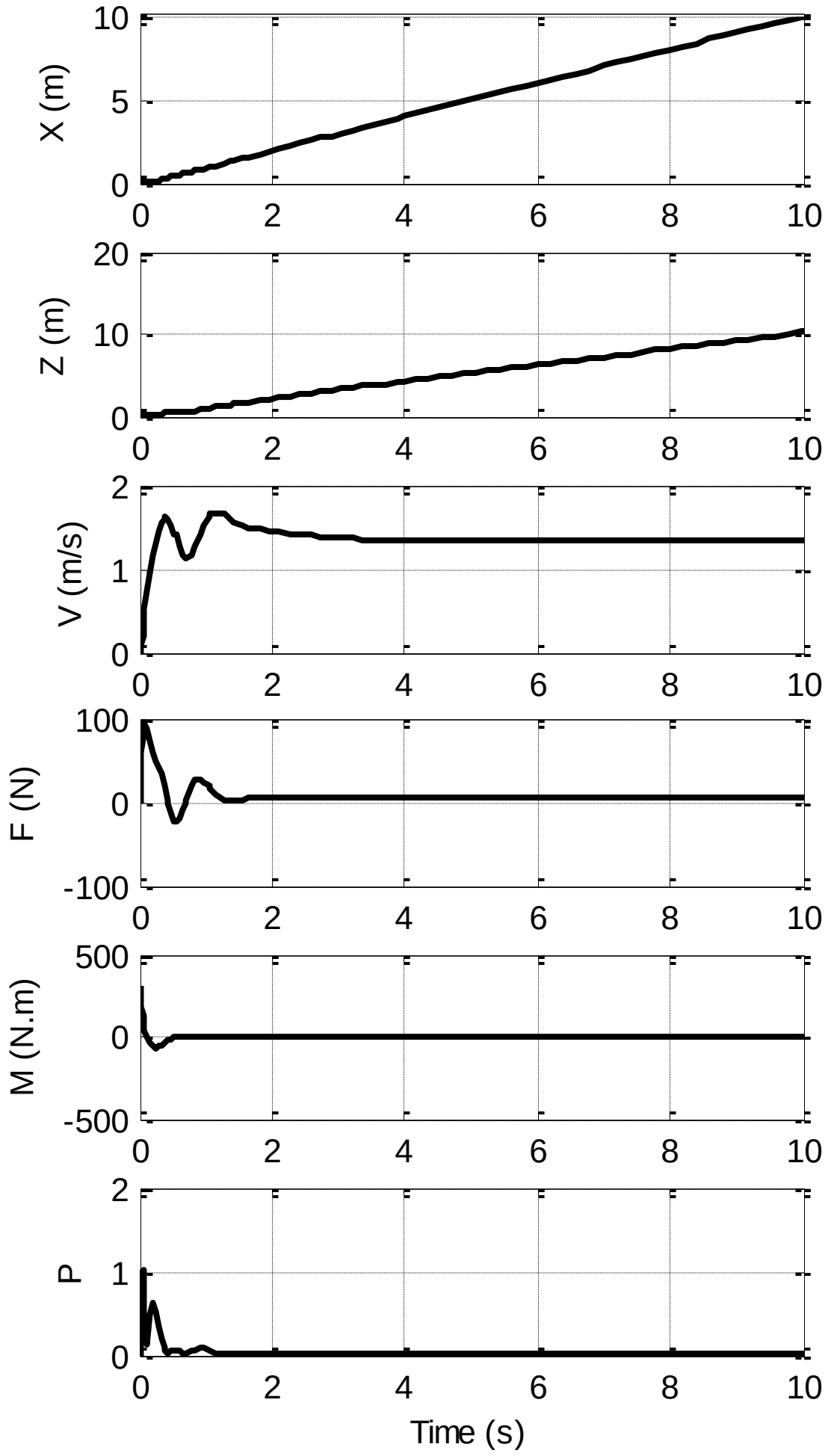

Figure 14. $x_I$, $z_I$, velocity, force, moment, and nondimensional power in response to reference inputs of $x_{ref}$ and $z_{ref}$ both as ramp functions with slopes of 1.

The reference inputs in Figure 15 are ramp functions with a slope of 5 for $x_{ref}$, and Z2, given in Figure 12, for $z_{ref}$. The reference inputs in Figure 16 are a ramp function with a slope of 5 for $x_{ref}$, and Z3, given in Figure 12, for $z_{ref}$.

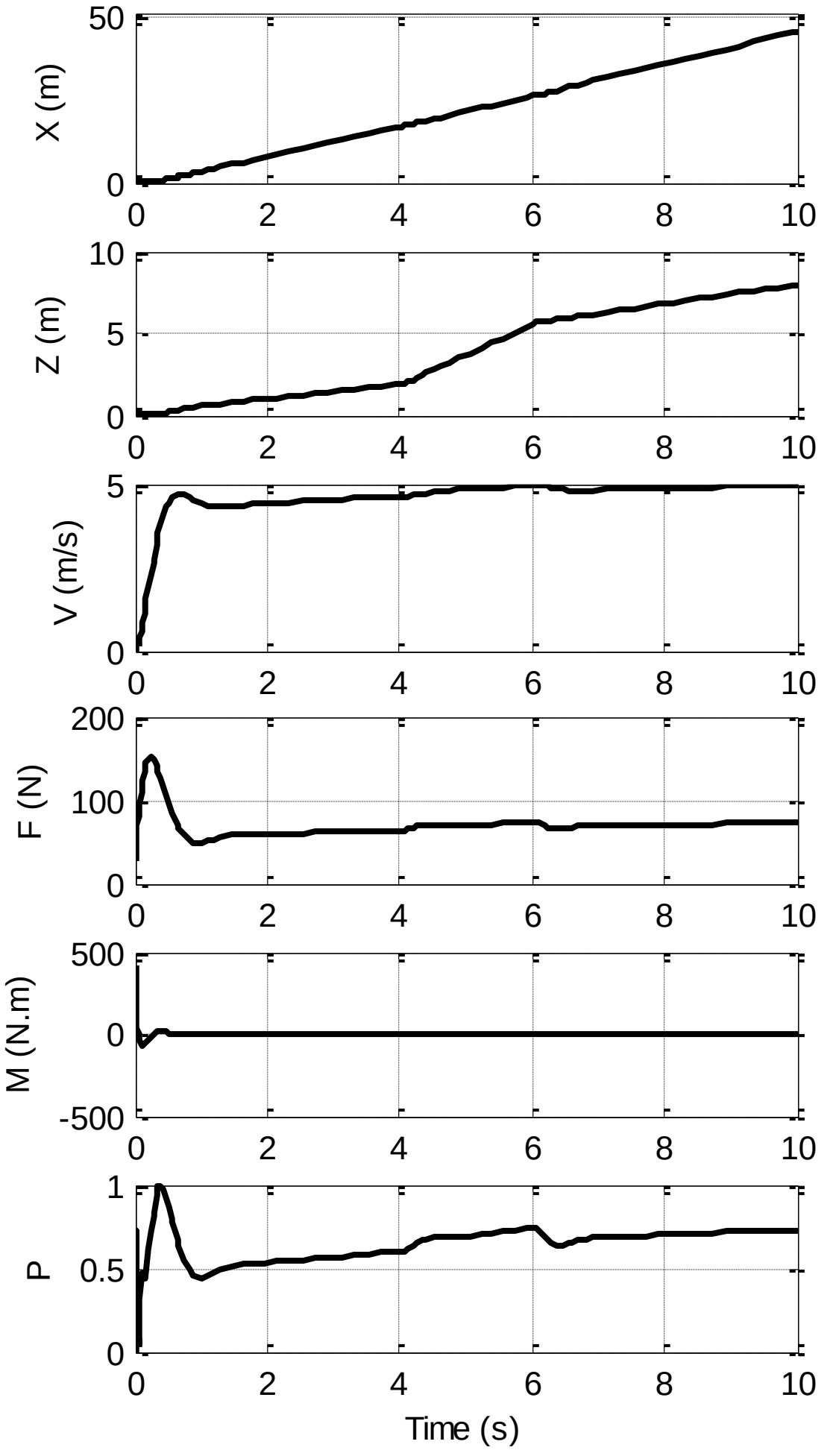

Figure 15. $x_I$, $z_I$, velocity, force, moment, and nondimensional power in response to reference inputs of $x_{ref}$ as a ramp function with a slope of 5, and $z_{ref}$ as Z2.

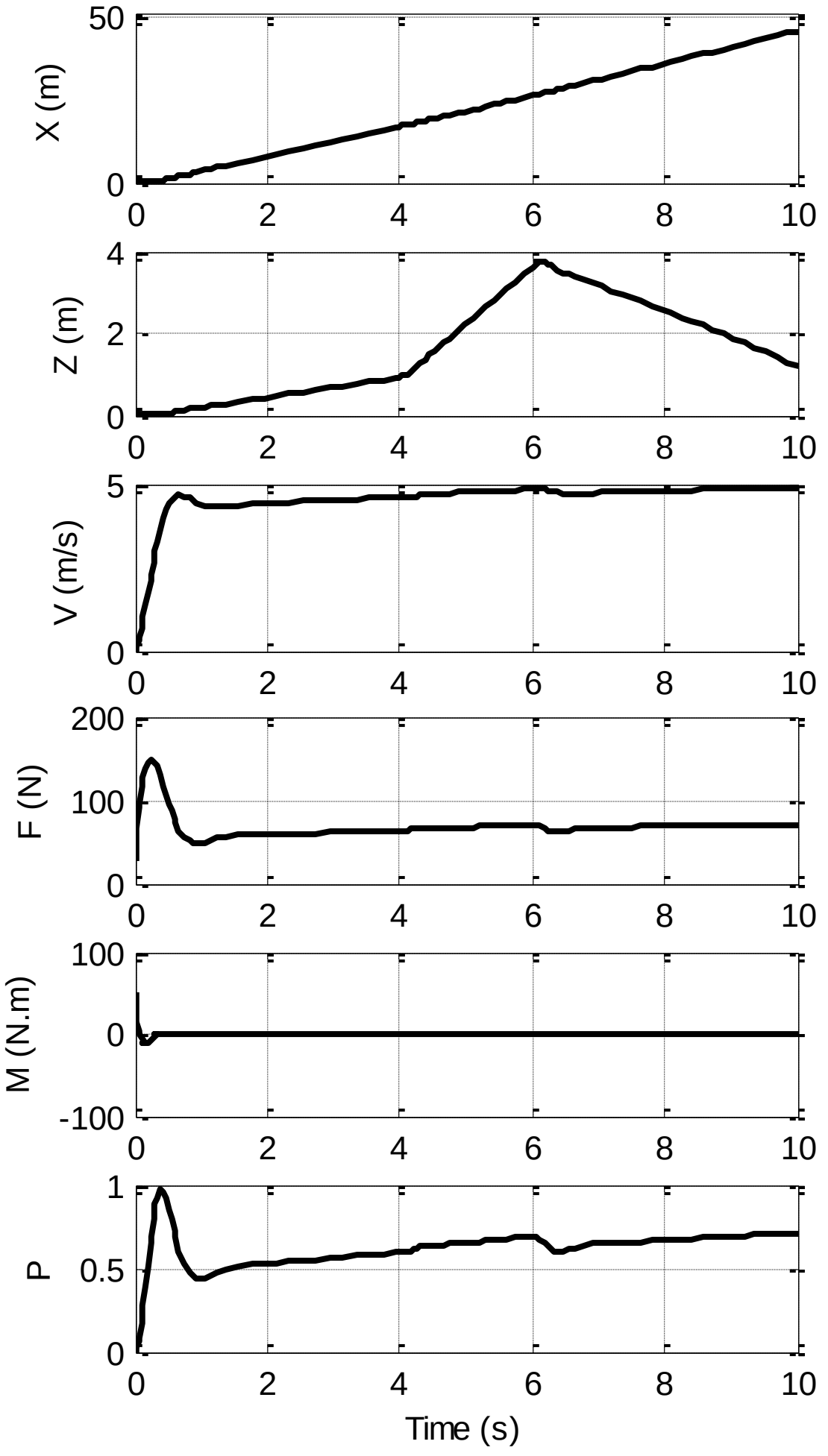

Figure 16. $x_I$, $z_I$, velocity, force, moment, and nondimensional power in response to reference inputs of $x_{ref}$ as a ramp function with a slope of 5, and $z_{ref}$ as Z3.

The reference inputs in Figure 17 are a ramp function with a slope of 5 for $x_{ref}$, and Z4, given in Figure 12, for $z_{ref}$. The reference inputs in Figure 18 are a ramp function with a slope of 5 for $x_{ref}$, and Z5, given in Figure 12, for $z_{ref}$.

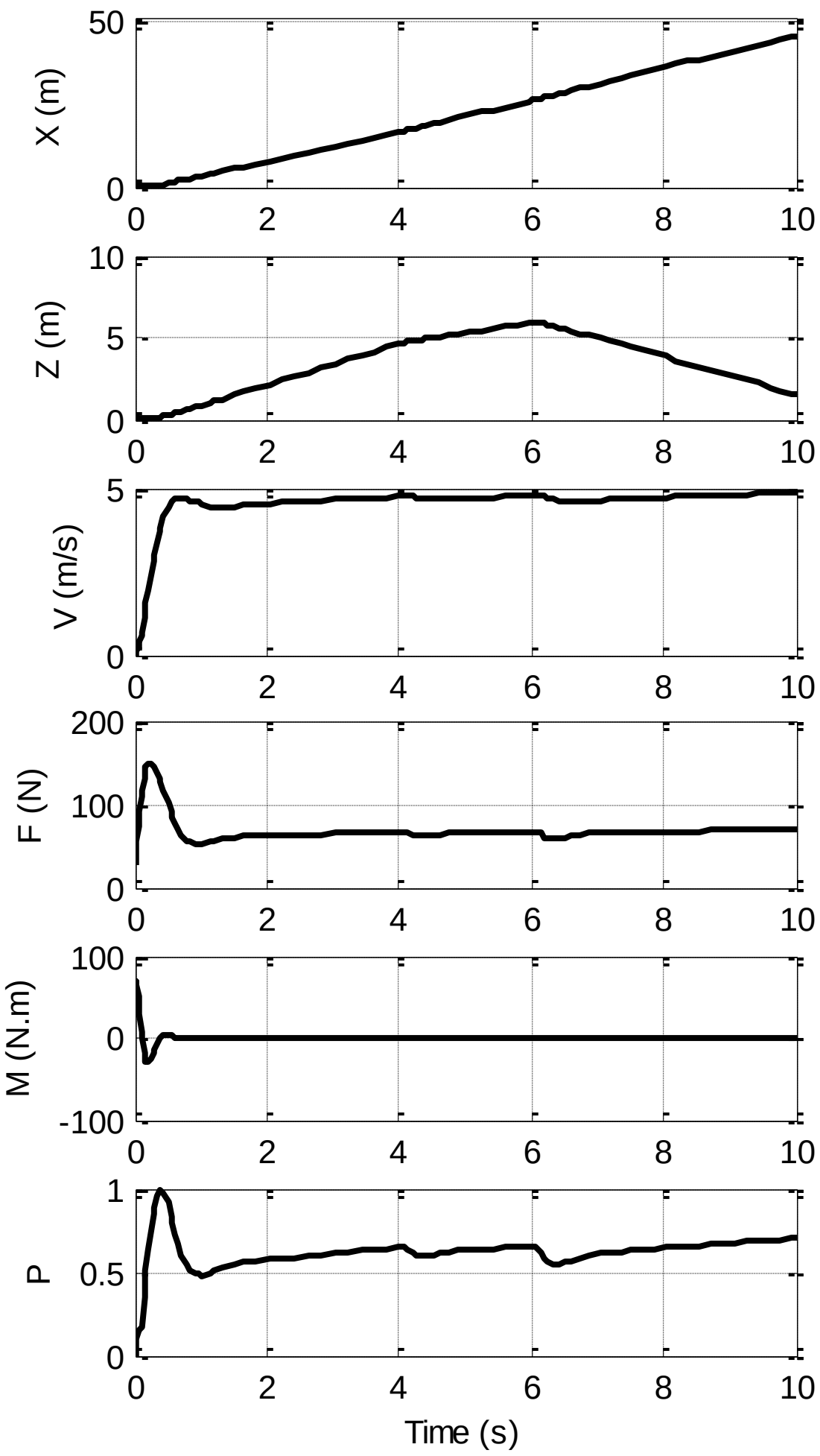

Figure 17. $x_I$, $z_I$, velocity, force, moment, and nondimensional power in response to reference inputs of $x_{ref}$ as a ramp with a slope of 5, and $z_{ref}$ as Z4.

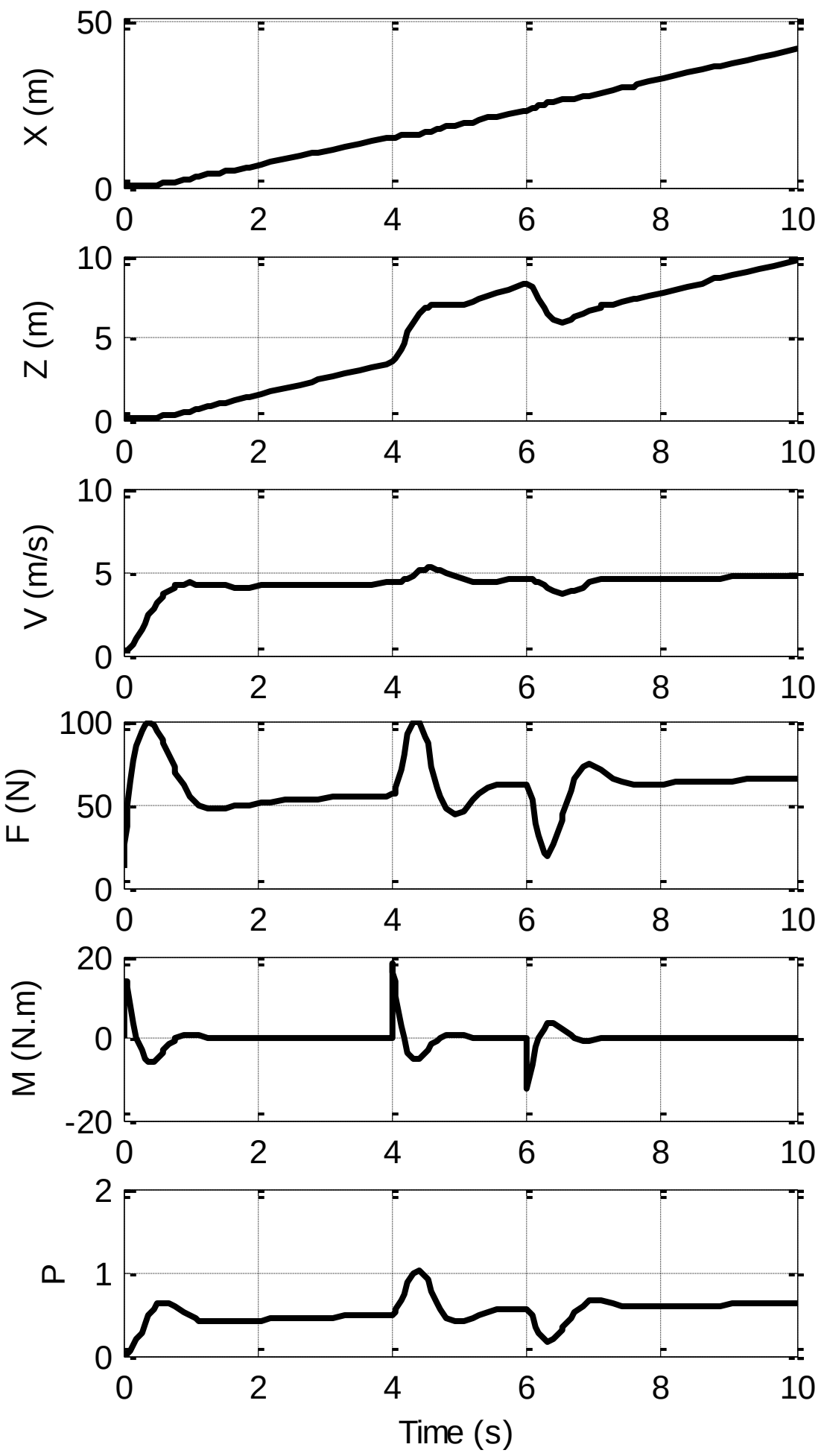

Figure 18. $x_I$, $z_I$, velocity, force, moment, and nondimensional power in response to reference inputs for $x_{ref}$ as a ramp function with a slope of 5, and $z_{ref}$ as Z5.

One of the constraints on the trajectory tracking problem is $P_{non} \leq 1$, or self-powered condition, for each duty cycle in Figure 12. Self-powered trajectory tracking has been achieved in all examples presented in Figure 13 to Figure 18. However, it is still required that other constraints in (30) to be satisfied. For instance, if the trajectory of the UAV is found as $\boldsymbol{r}$, or in scalar form as $x_I$ and $z_I$, and the constrained lower and upper bounds are $\boldsymbol{r} \in \boldsymbol{\mathcal{X}}_2 := [\boldsymbol{r}_{min}, \boldsymbol{r}_{max}]$ m, or in scholar form as the $x_I \in \boldsymbol{\mathcal{X}}_2 := [x_{I_{min}}, x_{I_{max}}]$ m, and $z_I \in \boldsymbol{\mathcal{X}}_2 := [z_{I_{min}}, z_{I_{max}}]$ m.

Due to various sources of uncertainties such as wind disturbances, available solar power, etc., the vehicle response can be uncertain, and therefore the lower and upper bounds ensure that the self-powered condition is attained while the performance parameters remain in the acceptable range. Therefore, if the UAV path (e.g., $\boldsymbol{r}$) fails to remain in the lower and upper bound of the desired trajectory, the self-powered trajectory tracking scheme is not achieved. Let us now assume that the trajectory obtained in Figure 18 in tracking $z_{ref}$ (i.e., Z5 in Figure 12) is not in the span of the desired lower and upper bounds (i.e., $z_I \leq z_{I_{min}}$ or $z_I \geq z_{I_{max}}$). Thus, larger actuator forces and moments are required to maintain the trajectory tracking error in the acceptable region. The results in Figure 19 illustrate improved trajectory tracking (i.e., smaller error) which is associated with larger actuator forces and moments, as shown in the figure. This leads to power demand up to 50 times larger than the available solar power. Therefore, the more power, which also corresponds to faster actuation response, leads to failing of the self-powered capability in this example. Further discussion on the response associated with the velocity and acceleration constraints in (30) is presented in the next section. Figure 20 focuses on time duration of 3 seconds to 7 seconds, which includes to the highest power demand at around 4 seconds time. Figure 21 and Figure 22 are presented to discuss the solar-powered speed introduced in Section VI (e.g. Figure 9 and Figure 10), in a more general flight condition (rather than the constant speed) such as the response to a step input. The vehicle experiences almost a constant velocity of about 5.5 m/s for a short period of time. During this constant velocity the nondimensional power is almost equal to 1. The small deviation of the velocity and power values (in comparison with the solar-powered speed in Section VI) is due to the additional power consumed to accelerate the vehicle (although small acceleration during this time). Thus, this velocity agrees with the solar-powered speed in Figure 9 with 10% PV efficiency and L/D=1.

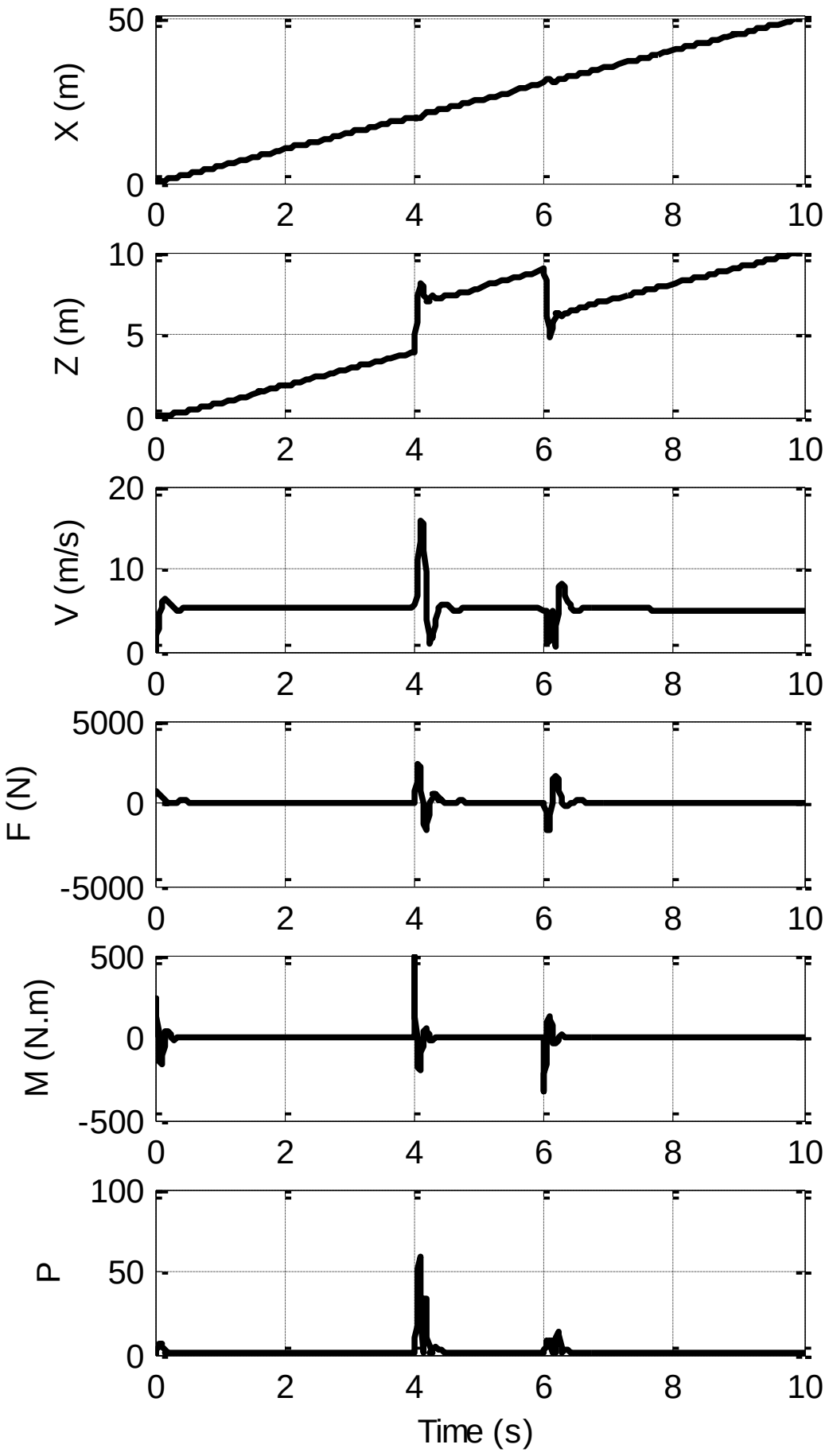

Figure 19. Results associated with the fast response; $x_I$, $z_I$, velocity, force, moment, and nondimensional power in response to reference inputs with $x_{ref}$ as a ramp with a slope of 5, and $z_{ref}$ as Z5.

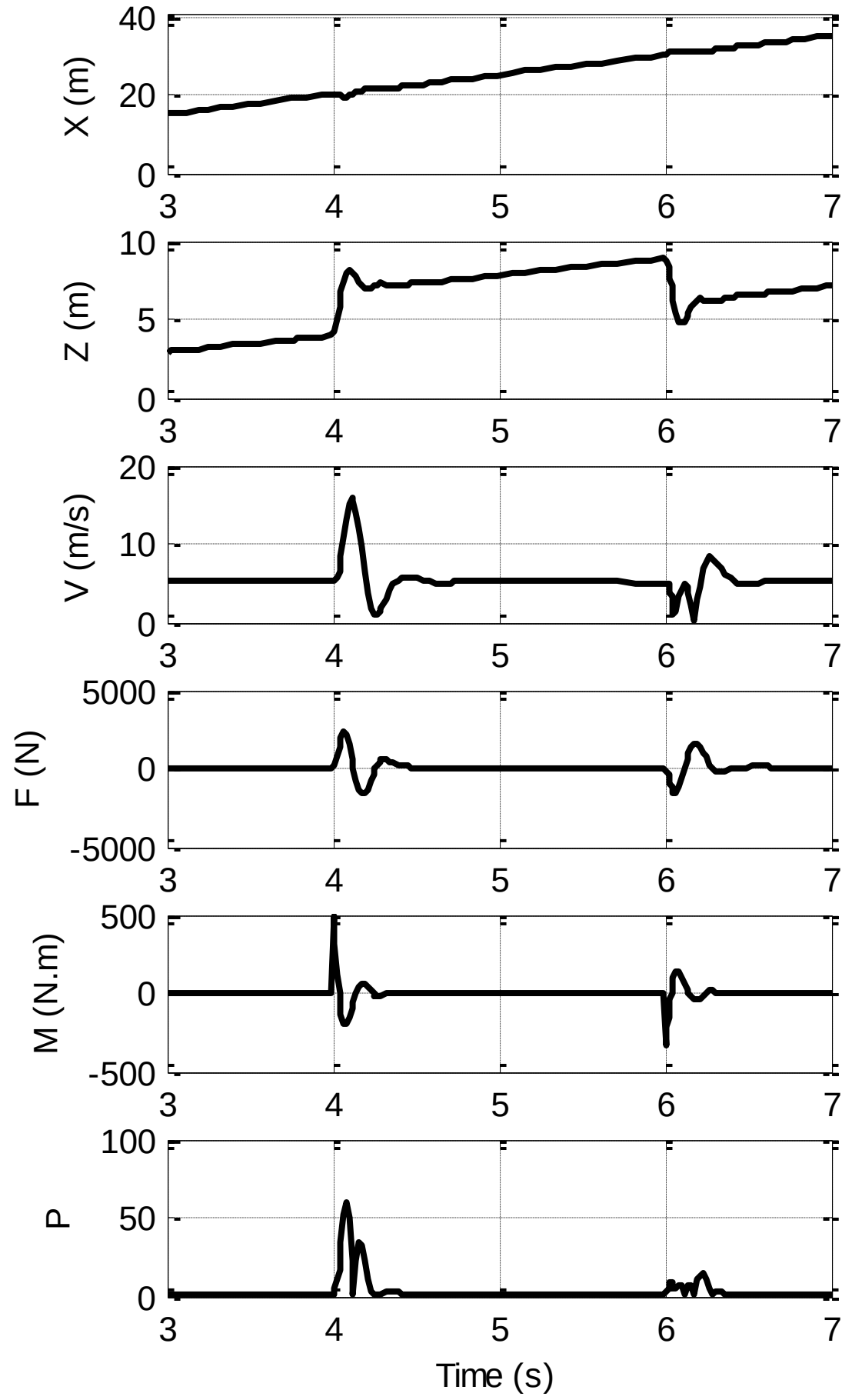

Figure 20. Results associated with the fast response; $x_I$, $z_I$, velocity, force, moment, and nondimensional power in response to reference inputs with $x_{ref}$ as a ramp with a slope of 5, and $z_{ref}$ as Z5.

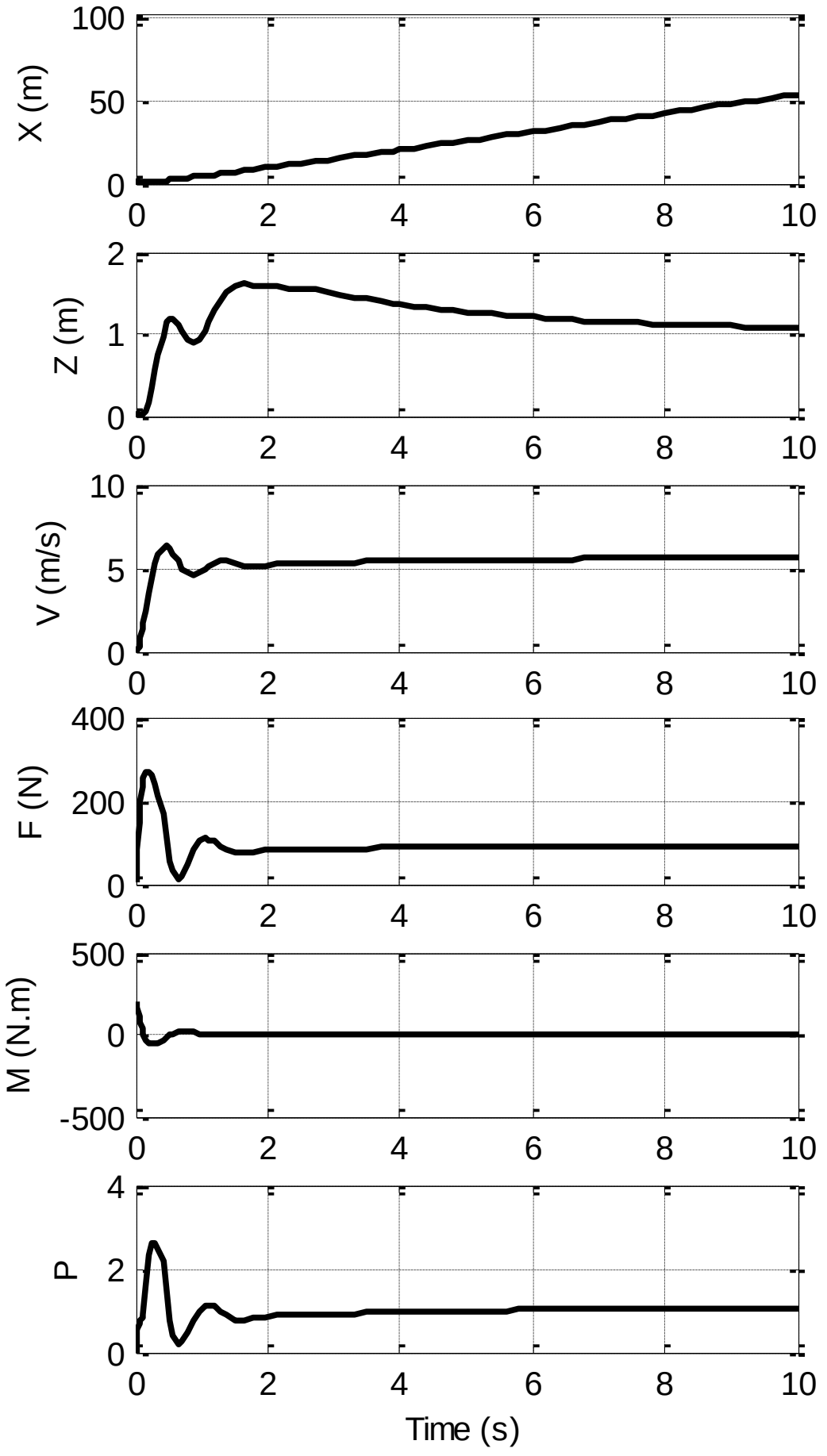

Figure 21. Demonstration of solar-powered speed in response to a step input (at 5.5 m/s velocity).

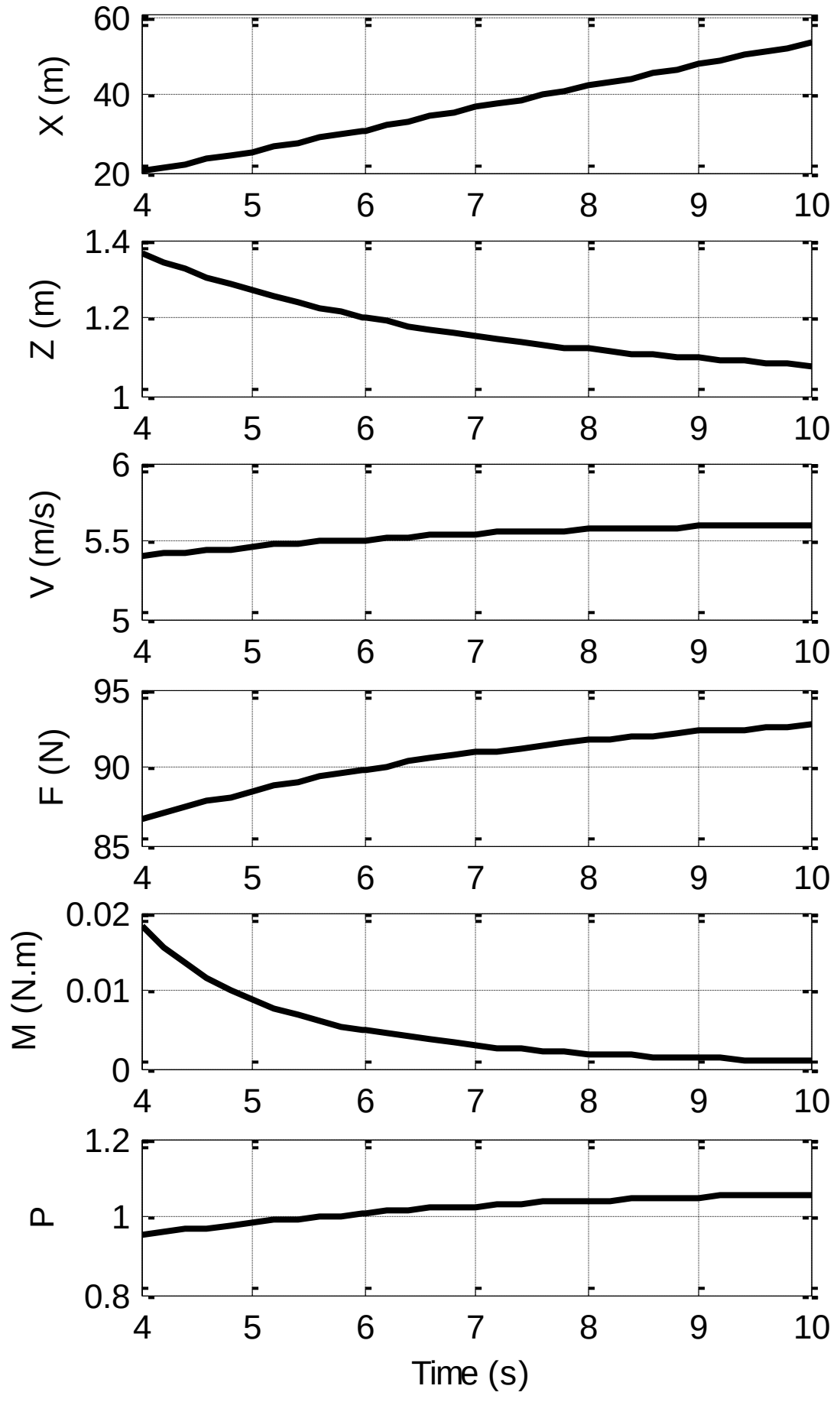

Figure 22. Demonstration of solar-powered speed in response to a step input (at 5.5 m/s velocity).

Analysis of the self-powered scheme by considering upper and lower bounds for the controller gains, velocity, overshoot, peak time are discussed in the next section.

## VIII. Self-powered dynamics and control

The self-powered control scheme is presented here as a function of controller parameters. Lower and upper bounds for displacement, velocity and acceleration are considered. Equation (1) gives the dynamics of the UAV, and Figure 7 represents the feedback control system in this problem. The nondimensional power, $P_{non}$, is obtained as a function of proportional and derivative controller gains associated with propeller thrust of motor $i$, with the lower and upper bounds as

$$k_{P,F,i} \in \mathcal{X}_{P,F,i} := \left[k_{P,F,i,min}, k_{P,F,i,max}\right], k_{D,F,i} \in \mathcal{X}_{D,F,i} := \left[k_{D,F,i,min}, k_{D,F,i,max}\right]$$

where $\mathcal{X}_{P,F,i}$, $\mathcal{X}_{D,F,i}$ indicate that the nondimensional power is a function of the stochastic proportional gain, $k_{P,F,i}$, and derivative gain, $k_{D,F,i}$, respectively, in the optimization process. The upper and lower bounds represent the acceptable range of stochastic or incomplete information in regards with each of the parameters, and should result in the probability of $\mathbb{P}[P_{non}(X) > 1] \leq 0$ (i.e., $P_{non} \leq 1$) corresponding to the state that the self-powered scheme does not fail. If the UAV control performance specifications fail to remain in the lower and upper bound range, the self-powered control scheme is not achieved.

In a one-dimensional problem, a resultant thrust force acting along one direction can be assumed, and therefore an equivalent motor and propeller ($i$=1) which gives this thrust is taken into account. $P_{non} = \frac{P_c}{P_g}$ is calculated based on the UAV parameters in Section VII. $P_c$ is given by Equation (6), and $P_g$ is determined based on PV cell efficiency of 10%. The feedback control problem is solved for each combination of controller gains, $k_{P,F,1}$ and $k_{D,F,1}$, and maximum consumed power for $P_c$, associated with each gain combination, is determined. The result for $P_{non}$ in this problem is obtained for $k_{P,F,1} \in \mathcal{X}_{P,F,1} := [0, 1000]$, $k_{D,F,1} \in \mathcal{X}_{D,F,1} := [0, 1000]$. The result for nondimensional power is shown in Figure 23.

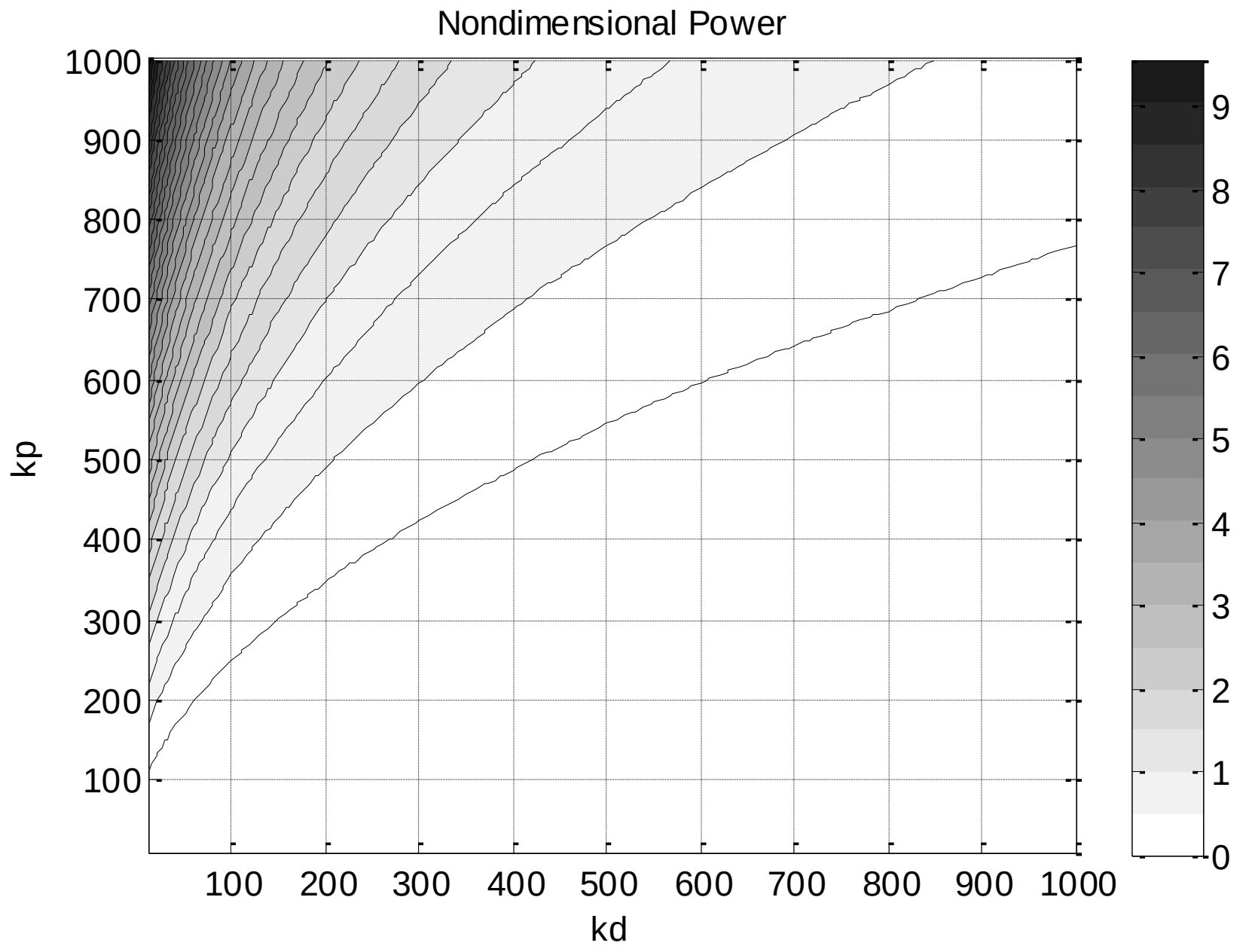

Figure 23. $\boldsymbol{P_{non}}$ as a function of controller proportional and derivative gains.

The region corresponding to nondimensional power $P_{non} \leq 1$ in Figure 23 gives the desired controller gains, $k_{P,F,1}$ and $k_{D,F,1}$, which results in self-powered control of the system. It is now required to investigate the controller performance in terms of the desired displacement, velocity and acceleration as discussed below. The solution of the feedback controller in Figure 7, and Equation (1) for the dynamics of the UAV, is applied here for analyzing the vehicle. The solution is repeated for every combination of $k_{P,F,1}$ and $k_{D,F,1}$ for the entire range of these control gains, and upper bounds for displacement, velocity and acceleration results are found as follows. The desired range for each kinematic parameter, which is within the desired upper and lower bound range, is obtained for $P_{non} \leq 1$.

The desired displacement lower and upper bounds can be

$$\boldsymbol{r} \in \boldsymbol{\mathcal{X}}_2 := [\boldsymbol{r}_{min}, \boldsymbol{r}_{max}] \text{ m}$$

In a two-dimensional problem, the desired upper and lower bounds are defined in scalar forms as $x_I := [x_{I_{min}}, x_{I_{max}}]$ m, and $z_I := [z_{I_{min}}, z_{I_{max}}]$ (or $x_{I_{min}} \leq x_I \leq x_{I_{max}}$, and $z_{I_{min}} \leq z_I \leq z_{I_{max}}$). In this one-dimensional problem, only $x_I$ (or $z_I$) component is under investigation. The reference input is a unit step function here, and the controller performance specification is set to the maximum overshoot of 1.6 m. Therefore, the upper bound is 1.6 m, or $x_I := [1, 1.6]$ m (or $z_I := [1, 1.6]$ m). For a fully buoyant vehicle, $z_I$ and $x_I$ results are analogues when solving a one-dimensional problem in either direction. The upper bound represents the overshoot in this example, and not the displacement steady-state error. $x_I$ is plotted in Figure 24.

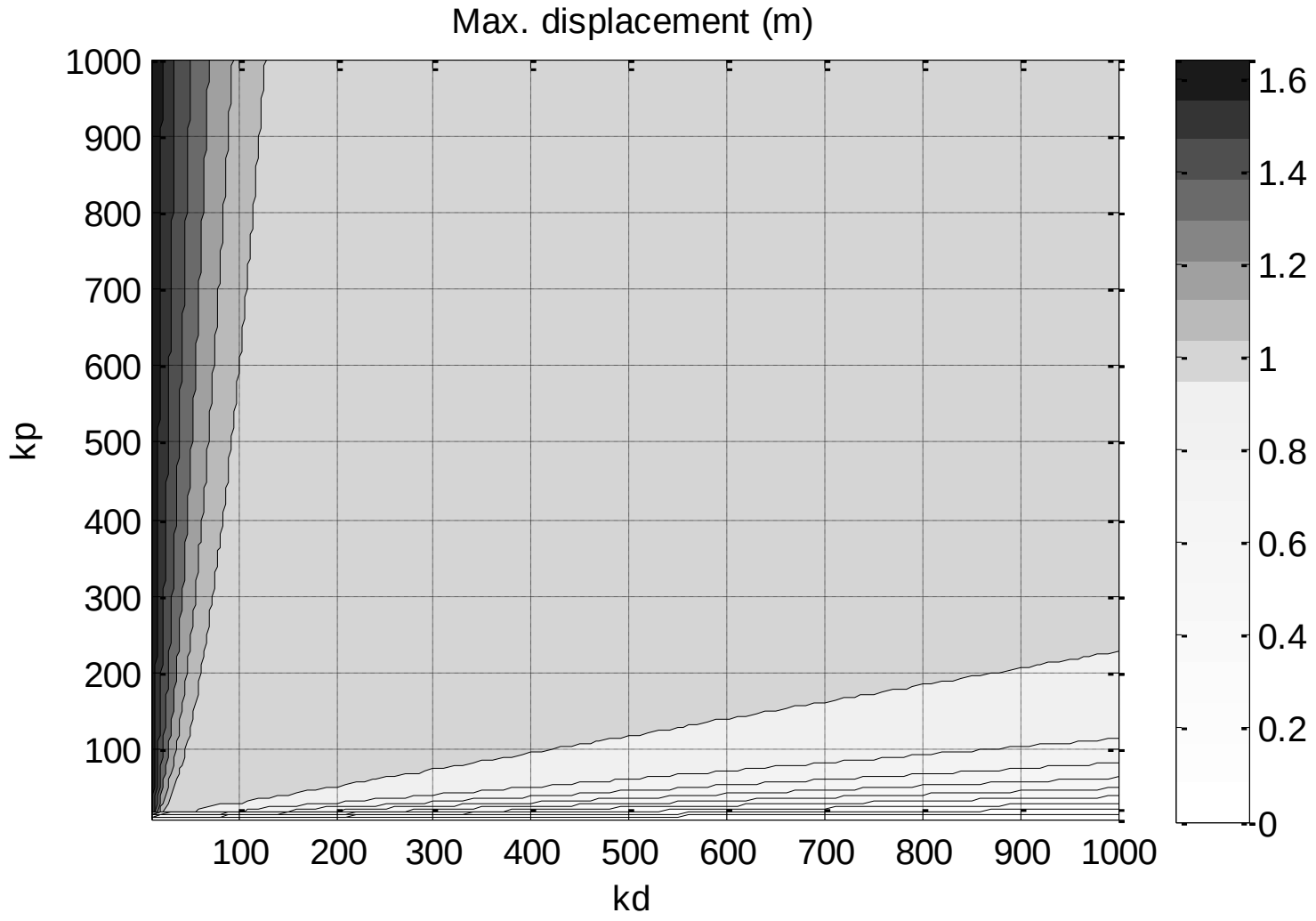

Figure 24. Upper bound, or overshoot, for displacement $x_I$, as a function of controller proportional and derivative gains.

The velocity of the vehicle ($\dot{x}_I$, in a one-dimensional problem) is another factor in specifying the control performance. The feedback control problem in Figure 7, in conjunction with Equation (1) which identifies the dynamics of the vehicle, is solved and maximum speed for each combination of $k_{P,F,1}$ and $k_{D,F,1}$ is obtained. The generalized upper and lower bound velocity vector can be assumed as

$$\boldsymbol{v_B} \in \boldsymbol{\mathcal{X}}_3 := [\boldsymbol{v_B}_{min}, \boldsymbol{v_B}_{max}] \text{ m}/s$$

These velocity upper bounds are shown in Figure 25.

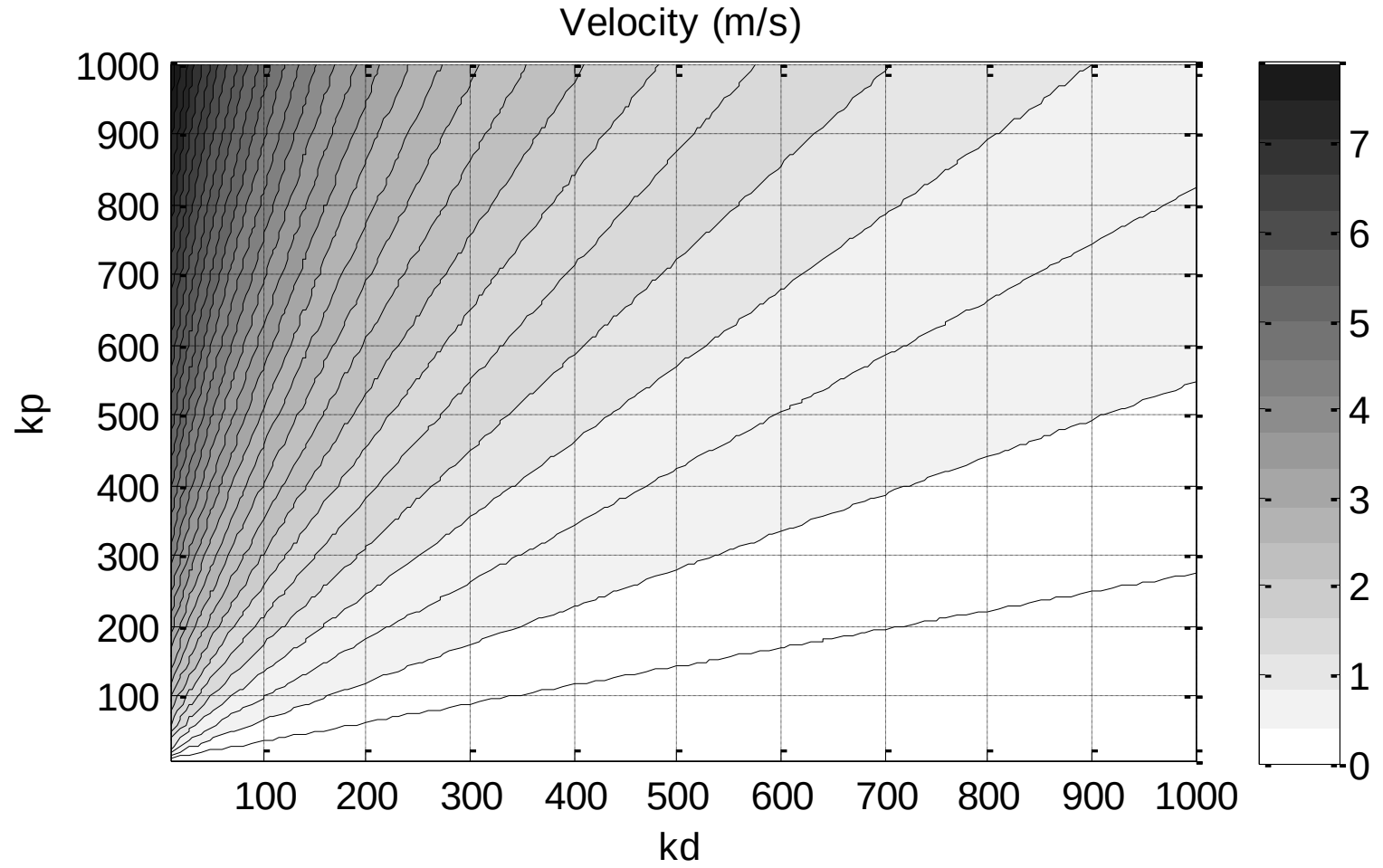

Figure 25. Upper bound of velocity $\dot{x}_I$, as a function of controller proportional and derivative gains.

In this problem, the velocity lower bound can be set to avoid large time delays in the response. For instance, $\dot{x}_I \geq 1\ \text{m}/\text{s}$ may be set as the speed lower bound, according to the desired performance specification. It is required to obtain the overshoot and velocity lower and upper bounds by referring to Figure 24 and Figure 25, respectively, in addition to the self-powered condition in Figure 23.

The acceleration range may be investigated by evaluating the peak time as the next control performance specification. Again, the feedback control problem in Figure 7, in conjunction with Equation (1) is solved and peak time for each combination of $k_{P,F,1}$ and $k_{D,F,1}$ is obtained. The results are presented in Figure 26.

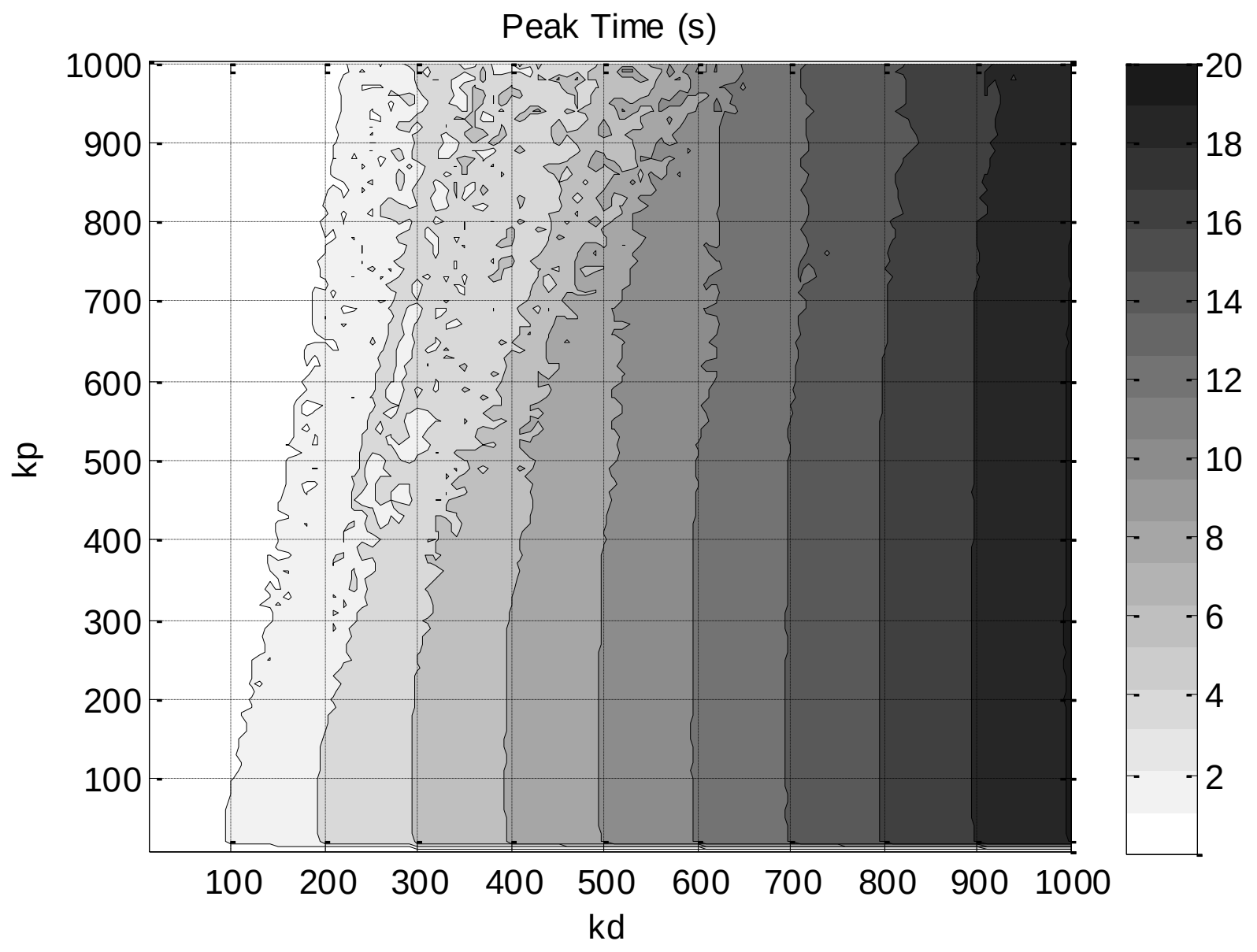

Figure 26. Upper bound of velocity $\dot{x}_I$, as a function of controller proportional and derivative gains.

By refereeing to Figure 26, and based on performance specification the region with large values of peak time is avoided, depending on design requirements, while taking into account the self-powered condition.

Other parameters with lower bounds and upper bounds may be considered in solving a problem, if required.

As a summary of the results in this section, an example of selecting suitable values of controller gains is given as follows. Higher values of $k_{D,F,1}$ is desired for achieving $P_{non} \leq 1$ (Figure 23), and smaller overshoot (Figure 24). Larger values of $k_{P,F,1}$ is desired for achieving faster response or larger velocity (Figure 25) and smaller peak time (Figure 26). Therefore, suitable gains can be estimated, for instance as $k_{P,F,1}$ = 100 and $k_{D,F,1}$ = 100, where $P_{non} \leq 1$, while attaining satisfactory performance specifications as: no overshoot, and velocity and peak time values equal to 1.

## IX. Conclusion

This paper explored the conditions and scenarios for developing Self-powered Solar Unmanned Aerial Vehicles (UAVs) as energy independent vehicles in accordance to dynamics, control system and solar energy accessible to the vehicles. Novel solar-powered multi-rotor electric aerial vehicles were discussed in this paper. These aerial vehicles include quadrotor, trirotor, and octorotor configurations, which take advantage of the buoyancy force for lift, and solar energy for the needed electrical power. A scheme for self-powering was explored in achieving long-endurance operation, with the use of solar power and buoyancy lift. The ultimate goal has been the capability of "infinite" endurance while giving proper consideration to the dynamics, control performance, maneuvering, and duty cycles of UAVs. Nondimensional power terms were obtained in association with the UAV power demand and solar energy input, in a framework of Optimal Uncertainty Quantification (OUQ). Solar-powered speed was introduced for cuboid and ellipsoid hull geometries. The corresponding plots of solar-powered speed versus nondimensional geometry of the vehicle introduced as a generalized reference for achieving the speed powered by solar energy in real time. Controller gain values were obtained for attaining self-powered operation associated with various trajectories. The problem of self-powered dynamics and control was discussed by obtaining lower and upper bounds for overshoot, velocity, and peak time in terms of controller gains. The discussed class of aerial vehicles can overcome the limited flight time of current electric Unmanned Aerial Vehicles, thereby expanding such application domains as aerial robotics, monitoring and inspection, safety and security, search and rescue, and transport. Extended various duty cycles, and partially buoyant vehicles for specific applications, and various disturbance conditions, design configurations, and control strategies, are considered for future investigations.

## Acknowledgments

The first author wishes to acknowledge Brunel Research and Innovation Fund Award for support of this research. Jason D. Rhodes, Alina A. Kiessling, and Marco B. Quadrelli are supported by Jet Propulsion Laboratory, which is operated by the California Institute of Technology under a contract with the National Aeronautics and Space Administration (NASA).

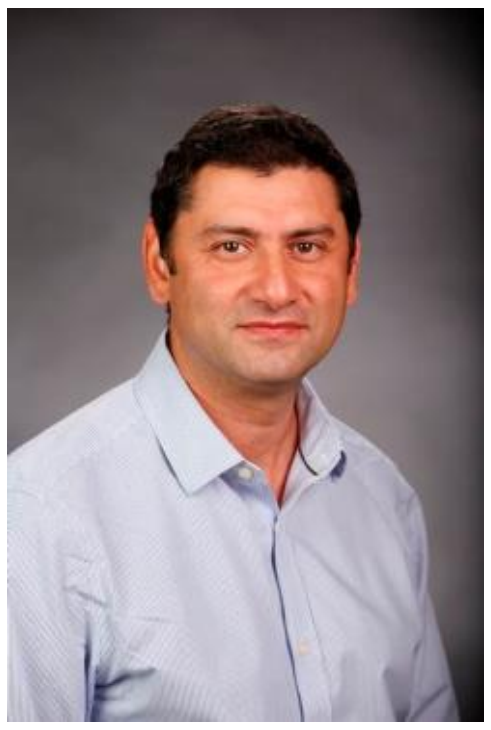

**Farbod Khoshnoud**, PhD, CEng, PGCE, HEA Fellow, is a faculty at California State Polytechnic University, Pomona. His current research areas include Self-powered and Bio-inspired Dynamic Systems; Quantum Multibody Dynamics, Robotics, Controls and Autonomy, by experimental Quantum Entanglement, and Quantum Cryptography; Theoretical Quantum Control techniques. He was a research affiliate at NASA's Jet Propulsion Laboratory, Caltech in 2019; an Associate Professor of Mechanical Engineering at California State University; a visiting Associate Professor in the Department of Mechanical Engineering at the University of British Columbia (UBC); a Lecturer in the Department of Mechanical Engineering at Brunel University London; a senior lecturer at the University of Hertfordshire; a visiting scientist and postdoctoral researcher in the Department of Mechanical Engineering at UBC; a visiting researcher at California Institute of Technology; a Postdoctoral Research Fellow in the Department of Civil Engineering at UBC. He received his Ph.D. from Brunel University in 2005. He is an associate editor of the Journal of Mechatronic Systems and Control.

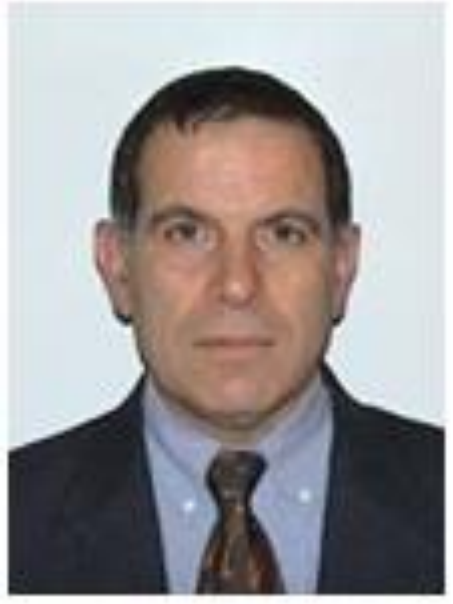

**Ibrahim I. Esat** has BSc and PhD from Queen Mary College of London University, after completing his Phd, he worked on mechanism synthesis at Newcastle university for several years, after this he moved to a newly formed University in Cyprus spending one and half year before returning back to UK joining to the University College of London University where he worked on designing "snake arm" robot. After this he moved to the Dunlop Technology division as a principal engineer working on developing CAD packages in particular surface modellers, following this he returned back to academia as a lecturer at Queen Mary College and later moving to Brunel University. He continued working closely with industry. He continued developing bespoke multi body dynamics software with user base in the UK, Europe and USA. Currently he is a full professor at Brunel University.

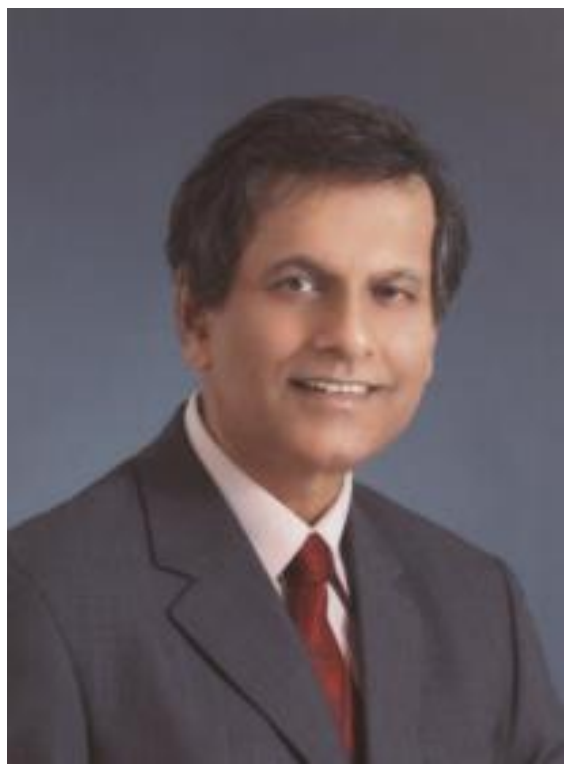

**Clarence W. de Silva** is a Fellow of: IEEE, ASME, Canadian Academy of Engineering, and Royal Society of Canada. He received Ph.D. degrees from Massachusetts Institute of Technology (1978); and University of Cambridge, U.K. (1998); and honorary D.Eng. degree from University of Waterloo, Canada (2008). He has been a Professor of Mechanical Engineering and Senior Canada Research Chair and NSERC-BC Packers Chair in Industrial Automation, at the University of British Columbia, Vancouver, Canada since 1988. He has authored 24 books and over 525 papers, approximately half of which are in journals. His recent books published by Taylor & Francis/CRC are: *Modeling of Dynamic Systems—with Engineering Applications* (2018); *Sensor Systems* (2017); *Senors and Actuators—Engineering System Instrumentation, 2nd edition* (2016); *Mechanics of Materials* (2014); *Mechatronics—A Foundation Course* (2010); *Modeling and Control of Engineering Systems* (2009); *VIBRATION—Fundamentals and Practice, 2nd Ed.* (2007); and by Addison Wesley: *Soft Computing and Intelligent Systems Design—Theory, Tools, and Applications* (with F. Karray, 2004).

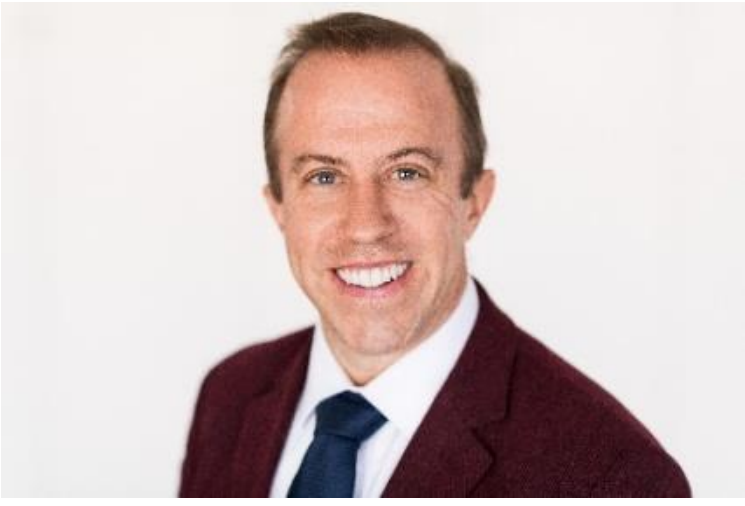

**Jason Rhodes** has a BS in Physics from Harvey Mudd College and a PhD in physics from Princeton. He worked as a postdoc at NASA GSFC and Caltech using Hubble Space Telescope data to study dark matter. He came to JPL in 2004, where he has worked on dark energy mission development for the past 14 years. He is the US science lead for ESA's Euclid mission, the JPL Project Scientist for NASA's Wide Field Infrared Survey Telescope (WFIRST). In addition to his research in dark energy, he has taken an interest in using coronagraphs and starshades to study the properties of exoplanets.

**Alina Kiessling** hails from Melbourne, Australia and has wanted to understand how the Universe works since she was in elementary school. She received her BS in Space Science from LaTrobe University in Australia, PhD in Astrophysics from Edinburgh University in Scotland, and completed a posdoc at JPL before becoming a research scientist there in 2014. Alina's background is in dark energy cosmology and she is a scientist on the European Space Agency's Euclid, NASA's Wide Field Infrared Survey Telescope (WFIRST), and the NSF/DOE Large Synoptic Survey Telescope (LSST). She is also the Deputy Study Scientist for the Habitable Exoplanet Observatory (HabEx) concept study, which has driven her on a steep learning curve into exoplanet science and mission formulation.

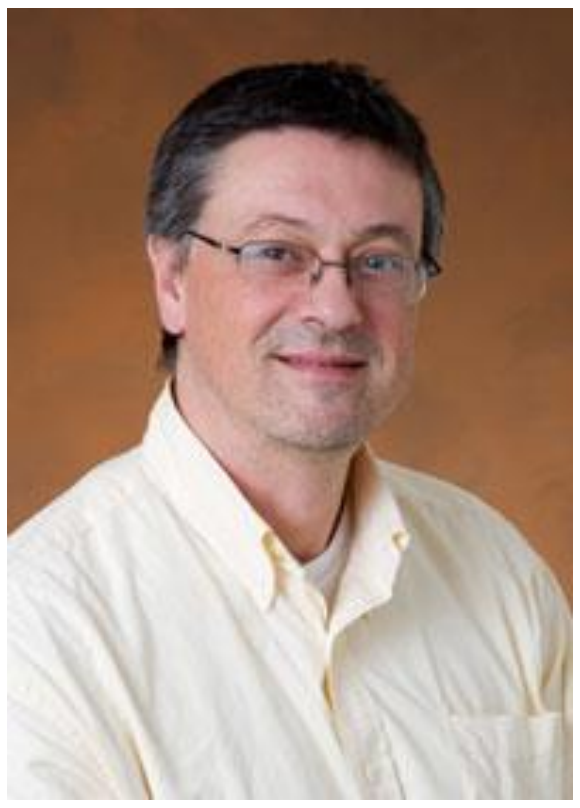

**Dr. Marco B. Quadrelli** is the supervisor of the Robotics Modeling and Simulation Group at the Jet Propulsion Laboratory (JPL), California Institute of Technology. He received a Laurea in mechanical engineering from the University of Padova, a M.S. in aeronautics and astronautics from MIT, and a Ph.D. in aerospace engineering from Georgia Institute of Technology in 1996. He has been a Visiting Scientist at the Harvard-Smithsonian Center for Astrophysics, did postdoctoral work in computational micromechanics at the Institute of Paper Science and Technology, and has been a Lecturer in aerospace engineering at both the JPL and at the California Institute of Technology Graduate Aeronautical Laboratories. He is a NASA NIAC Fellow, a Keck Institute for Space Studies Fellow, and an AIAA Associate Fellow. Dr. Quadrelli's flight project experience includes the Cassini–Huygens Probe Decelerator; Deep Space One; the Mars Aerobot Program; the Mars Exploration Rover and Mars Science Laboratory Entry, Descent, and Landing; the Space Interferometry Mission; and the Laser Interferometry Space Antenna. He has been involved in several research projects in the areas of tethered space systems;

distributed spacecraft and robots; active granular media; hypersonic entry and aeromaneuvering; planetary sampling; integrated modeling of space telescopes and inflatable spacecraft; and surface vessel dynamics and state estimation.